\documentclass[prd,amsmath,amssymb,reprint,superscriptaddress,floatfix,nofootinbib]
{revtex4-1}
\usepackage{amsmath,amssymb}
\usepackage{hyperref}
\usepackage{graphicx}
\usepackage[utf8]{inputenc}
\usepackage{bbm}
\usepackage{multirow}
\renewcommand{\Re}{\ensuremath{\operatorname{Re}}}
\newcommand{\tr}{\ensuremath{\operatorname{Tr}}}
\newcommand{\id}{\ensuremath{\mathbbm{1}}}
\newcommand{\betapf}[1]{\ensuremath{\beta^{-\frac{#1}{2}}}}
\newcommand{\dtau}{\ensuremath{\epsilon}}

\newcommand{\scft}{\ensuremath{\phi(x,\tau)}}
\newcommand{\SA}{\ensuremath{\mathcal{S}}}
\newcommand{\vev}[1]{\ensuremath{\left\langle#1\right\rangle}} 
\newcommand{\sev}[1]{\ensuremath{\left\langle#1\right\rangle}_{\eta}}
\newcommand{\sbev}[1]{\ensuremath{\left\langle#1\right\rangle}_{\bar{\eta}}}
\newcommand{\etab}{\ensuremath{\bar{\eta}}}
\newcommand{\order}[1]{\ensuremath{\mathcal{O}(#1)}}
\newcommand{\Ud}{\ensuremath{U^\dagger}}
\newcommand{\Mmu}{\ensuremath{M_\mu}}
\newcommand{\Mmud}{\ensuremath{M_\mu^\dagger}}

\newcommand{\nmeas}{\ensuremath{n_{\text{m}}}}
\newcommand{\nup}{\ensuremath{n_{\text{up}}}}
\newcommand{\ninit}{\ensuremath{n_{\text{init}}}}
\newcommand{\dash}{\textrm{--}}

\graphicspath{ {./plots/} {./} }

\begin{document}
\title{Universal renormalons in principal chiral models}
\newcommand{\RegensburgUniversity}{Institut f\"ur Theoretische Physik,
  Universit\"at Regensburg, 93040 Regensburg, Germany}
\newcommand{\FHJoanneum}{Institut Informationsmanagement, FH JOANNEUM, Eckertstraße 30i, 8020 Graz, Austria}
\author{Falk Bruckmann}
\affiliation{\RegensburgUniversity}
\author{Matthias Puhr}
\affiliation{\RegensburgUniversity}
\affiliation{\FHJoanneum}

\begin{abstract}
Perturbative expansions in many physical systems yield ``only'' asymptotic series which are
not even Borel resummable. Interestingly, the corresponding ambiguities point to
nonperturbative physics. We numerically verify this renormalon mechanism for the first
time in two-dimensional sigma models, that, like four-dimensional gauge theories, are
asymptotically free and generate a strong scale through dimensional transmutation. We
perturbatively expand the energy through a numerical version of stochastic quantization.
In contrast to the first energy coefficients, the high-order coefficients are independent
on the rank of the model. Technically, they require a sophisticated analysis of finite
volume effects and the continuum limit of the discretized model. Although the individual
coefficients do not grow factorially (yet), but rather decrease strongly, the ratios of
consecutive coefficients clearly obey the renormalon asymptotics.
\end{abstract}
%
%
\maketitle

Perturbation theory, the expansion in a small parameter, is a straightforward approach to
many physical systems, both classical and quantum. The high-order behavior of the
perturbative expansion, however, can be very elaborate: it may not be a convergent, rather
an {\em asymptotic series}. This situation already occurs for relatively simple quantum
mechanical systems, like the anharmonic oscillator expanded around zero anharmonicity, and
even for ordinary integrals (as toy models for path integrals)~\cite{LeGuillou:1990nq}.
In quantum field theories (QFTs) the asymptotic nature of an expansion in powers of the
coupling constant $\alpha$ hints at a physical instability, as pointed out by Dyson for
quantum electrodynamics~\cite{Dyson:1952tj}.

The typical dependence of an observable $E$ on
$\alpha$ contains factorially growing coefficients,
\begin{align}
 E=\sum_{n=0}^\infty c_n \alpha^n
 \quad (\alpha\to 0)
 \quad \text{with } c_n \!\stackrel{n\to\infty}{\sim}\! \gamma^n
  n^\kappa  n!.
 \label{eq:first}
\end{align}
Straightforward minimization and the application of Stirling's formula reveal that the
summand with the smallest magnitude comes at order $ n \approx 1 / \gamma \alpha $ and its value
$\exp(-1/\gamma\alpha)$ is a proxy for the limited accuracy of the asymptotic expansion.
If the coefficients $c_n$ have an alternating sign, the series~\eqref{eq:first} can be
Borel resummed and this way a unique value can be assigned to the observable $E$.
For sign coherent series, on the other hand, Borel resummation comes with an (imaginary)
{\em ambiguity} proportional (in leading order) to $\exp(-1/\gamma\alpha)$. In practical
calculations this is often not an issue, thanks to the smallness of the expansion parameter
$\alpha$. However, in asymptotically free systems --- such as four-dimensional non-Abelian
gauge and two-dimensional nonlinear sigma models --- the long-range regime is always
strongly coupled and the ambiguity can severely limit the ability to make physically
meaningful statements about $E$. These effects go under the name of {\em renormalons}
\cite{Beneke:1998ui}. Operator product expansion (OPE) offers another view on them as it
involves nonperturbative condensates and strong scales with a similar dependence on
$\alpha$ (see, e.g., \cite{Beneke:1998ui} and Eq.~\eqref{eq:gap} below).

From experience with various models, the emergence of sign-coherent asymptotic expansions
is connected to vacuum degeneracy, as labeled by topological quantum numbers. What makes
this subject so fascinating is that the {\em nonperturbative nature} of the perturbative
ambiguity seems to be connected to the system's nonperturbative classical tunneling
solutions with its typical factors of $\exp(-1/\alpha)$. This concerns instantons as
stable topological configurations and their superpositions, but also unstable saddles, as
is the case in the models at hand~\cite{Cherman:2014ofa}.  There is mounting evidence from
\emph{resurgence} theory calculations that, taken together, information from asymptotic
perturbative series and from topology leads to a cancellation of nonperturbative
ambiguities and could potentially be used to construct a nonperturbative continuum
formulation of
QFTs~\cite{Dunne:2015eaa,Dunne:2015ywa,Cherman:2013yfa,Dunne:2012zk,Dunne:2012ae}.

So far, a rigorous mathematical proof of the resurgence picture does not exist for most
theories of interest and the calculations rely on certain assumptions (or special features
of supersymmetric theories). While these assumptions are physically and mathematically
well motivated, it is nevertheless important to check their validity. Moreover, most
studies of the renormalon mechanism are based on the summation of a special class of
Feynman diagrams.
The only known \emph{ab initio} determination of high-order perturbative coefficients is provided through
numerical simulations on space-time lattices. The method of choice is not the common
Monte
Carlo framework, rather stochastic quantization (Langevin dynamics)
\cite{Parisi:1980ys} and the numerical version thereof, combined with numerical
perturbation theory to
{\em numerical stochastic perturbation theory} (NSPT)~\cite{DiRenzo:1994sy,DiRenzo:2004hhl}.
It has the great advantage that its effort grows only quadratically in the expansion order,
not factorially as in diagrammatic perturbation theory.
With this tool renormalons in four-dimensional $SU(3)$ Yang-Mills theory
have been clearly demonstrated in two observables: the lattice action
(plaquette) \cite{Bauer:2011ws,Bali:2014fea} and the energy of static sources
(Polyakov loop) \cite{Bali:2013pla}. High expansion orders (up to 35 in $1/\beta$),
extrapolations in volume and Langevin time, and other sophisticated lattice
methods had to be used to improve on previous studies that could not find
renormalons. Very recently, quarks have been included in that
framework~\cite{DelDebbio:2018ftu}.

Renormalons and resurgence in sigma models have been investigated before; see for example
\cite{Volin:2009wr, Dunne:2012ae, Dunne:2012zk, Cherman:2013yfa, Cherman:2014ofa,
  Fujimori:2016ljw, Fujimori:2018kqp, Ishikawa:2019tnw, Marino:2019eym}. 
There are however no NSPT studies of renormalons for theories other than quantum
chromodynamics. To provide cross-checks of the universality of the renormalon picture and
resurgence it is important to have reliable first-principle results for a large variety of
different theories. In this paper, we present first numerical results for the
perturbative coefficients of the energy density of 1+1 dimensional principal chiral models
$PC(N)$, with special emphasis on the universality in the rank $N$.  In these models, the
degrees of freedom are $SU(N)$ group valued fields, and the Euclidean action is nothing
but the obvious kinetic term
$\frac{1}{g^2}\int\!d^2x\,\text{tr}\,
\partial_\mu
U(x) \partial_\mu
U^\dagger(x)$.
Superficially this looks like the action of a free theory, but we emphasize that the
constraint $U \in SU(N)$ introduces couplings between the field components and makes the
$PC(N)$ models highly nontrivial. On lattices with spacing $a$ the derivatives translate
into nearest neigbor interactions, and the lattice action reads
\begin{align}
 \SA&=-\beta N\sum_{x,\mu} s_\mu(x)\,,
\label{eq:lat_action1} \\
 s_\mu(x)
 &=2\,\Re\tr\left( U(x)U^\dagger(x+a\hat{\mu}) \right), \quad U\in SU(N)\,.
\label{eq:lat_action2}
\end{align}
Like their $O(N)$~\footnote{Since $SU(2)$ is a three-sphere, there is an isomorphism between
  the models $PC(2)$ and $O(4)$.} and $CP(N-1)$ cousins, these sigma models are
asymptotically free~\footnote{Correspondingly, the continuum limit $a\to 0$ is achieved by
  $\beta \to \infty $.} and generate a mass and strong scale through quantum
fluctuations.

From a statistical physics analogy, the energy density $E$ is related to the
$\beta$-derivative of the partition function, in our convention
\begin{align}
a^2E
 =1-\frac{1}{4N^2V}\frac{\partial\log Z}{\partial\beta}
 =1-\frac{1}{4NV}\, \left\langle \sum_{x,\mu} s_\mu(x) \right\rangle,
\label{eq:plaquette_def}
\end{align}
where $V$ stands for the number of lattice sites. Note that
the energy density has mass dimension $d=2$.

As $\beta N$ is related to the inverse of $g^2$ [see Eq.~\eqref{eq:lat_action1}], both the
fields and the observable are expanded in powers of $\beta^{-1/2}$, 
\begin{align}
 U(x)
 &=\sum_{n=0}^\infty U_n(x) \beta^{-n/2}\qquad (\beta\to\infty)\,,\\
 a^2 E
 &=\sum_{n=0}^\infty E_n \beta^{-n}\,,
\end{align}
where we have immediately used that $E$ only contains integer powers of
$1/\beta$~\footnote{Note that in our approach the energy is formally expanded in
  half-integer powers, just like the fields, but every second of these
  coefficients is found to vanish, i.e., to be consistent with zero.}. Actually, the first few terms of this weak
coupling expansion are known analytically \cite{Brihaye:1983jr} and will be used as
benchmarks for our numerical results. For more details on the expansion within NSPT, we
refer to Appendix~\ref{sec:NSPT}.

To develop an expectation for the renormalon behavior of the energy expansion, we first of
all notice that, although their homotopy groups are trivial, $PC(N)$ models contain
nonperturbative saddles~\cite{uhlenbeck1989} (unitons), which may cure the ambiguity of a
sign coherent perturbative expansion~\cite{Cherman:2013yfa,Cherman:2014ofa}. Second, we
invoke from the large-$N$ expansion~\footnote{Where the 't Hooft coupling kept fixed is $g^2 N$ and thus
  $1/\beta $.} the two-loop relation between the
lattice spacing (inverse cutoff scale), the generated strong scale $\Lambda_L$
(proportional to the mass), 
and the bare lattice coupling $\beta$,\footnote{The coefficients $\beta _{0,1}$ parametrize the beta
  function of the coupling $\alpha = \beta ^{-1}$ with the cutoff: $a^{-1}\partial
  _{a^{-1}} \alpha = -\beta_0 \alpha^{2}-\beta_1\alpha^{3}+\protect
  \mathcal {O}(\alpha^{4})$; see e.g.,\ \cite
  {Rossi:1993zc,Gonzalez-Arroyo:2018aus}}
\begin{align}
 a\Lambda_L
 &=\sqrt{8\pi\beta}\, \exp(-8\pi\beta)\qquad (\beta\to\infty)
 \label{eq:gap}\\
 &=(\beta/\beta_0)^{\beta_1/\beta_0^2}\exp(-\beta/\beta_0)\,,\\
 &\quad\: \beta_0=\frac{1}{8\pi}\,,\quad
 \beta_1=\frac{1}{128\pi^2}\,.
\end{align}
The OPE relates the constant $\gamma$ in Eq.~\eqref{eq:first} to the running of the
coupling $\beta = 1/\alpha $ and the energy dimension of the observable $E$: since our observable has
mass dimension two, its perturbative part should, in leading order, receive
nonperturbative corrections of the form $\exp(-2\beta/\beta_0)$. We therefore anticipate
the energy coefficients to behave like
\begin{align}
E_n &\!\stackrel{n\to\infty}{\sim}\! \left(\frac{\beta_0}{2}\right)^n\!n!
\,,\qquad \frac{\beta_0}{2}=\frac{1}{16\pi}\,.  \label{eq:En_asympt}
\end{align}
That the energy possesses factorially growing sign coherent perturbative coefficients has
first been demonstrated in \cite{Fateev:1994ai}.
Applying the general arguments from below Eq.~\eqref{eq:first} to the expansion
coefficients $E_n$ themselves (e.g., by formally setting $\alpha=1$) using $\gamma = \beta_0/2$, the
$E_n$ are expected to start growing around order
\begin{align}
n^* \approx 16\pi \approx 50.
\label{eq:nstar}
\end{align}
Their ratios divided by the order,
\begin{align}
r_n:=\frac{E_n}{E_{n-1}\,n} &\!\stackrel{n\to\infty}{\sim}\!\frac{1}{16\pi}\,,  \label{eq:rn_asympt}
\end{align}
should approach a constant.
Note that these leading-order statements should hold independently of the rank $N$. Three-loop
corrections are related to the (regularization dependent) beta function coefficient $\beta_2$,
which contains order $\mathcal{O}(1/N^2)$ terms in the lattice scheme~\cite{Rossi:1993zc}.

We apply NSPT to calculate the expansion coefficients $E_n$ of the energy density on
symmetric two-dimensional lattices with the same number $L$ of sites in the spatial and
Euclidean time direction, $V=L \times L$, and periodic boundary conditions in both of
them. To study finite size effects and the assumed universality of
Eq.~\eqref{eq:rn_asympt}, we consider different lattice geometries and a variety of
different ranks $N$. In particular, we calculate the expansion coefficients up to $E_{10}$
for $N=12$ and even up to $E_{20}$ for $N=3,4,5,6$; see Appendix~\ref{sec:simdetails} for
details. The simulations for different $N$ are completely independent.

In Langevin simulations the
finite stochastic time step $\dtau$ introduces a systematic error and it is necessary to
perform the extrapolation $\dtau \to 0$. For the numerical integration of the Langevin
equation we utilize the Runge-Kutta algorithm from~\cite{Bali:2013pla}, which is exact up
to terms proportional to $\dtau^2$. We use up to five different values of $\dtau$ in our
simulations and perform the $\dtau \to 0$ limit by fitting a function linear in $\dtau^2$
to our data.

A remarkable feature of NSPT simulations is that neither the lattice spacing $a$ nor the
coupling $\beta$ enters the calculations explicitly. All computations are done directly with
the expansion coefficients. Therefore, it is not possible to assign a physical volume to
our lattice and it is not straightforward to go to the infinite
volume limit. Indeed, from OPE arguments very large finite size effects are
expected, even on the largest lattices that are achievable in present day
simulations~\cite{Bali:2013pla,Bali:2014fea}. Fortunately, the OPE enables us to infer the
functional form of the finite volume dependence, which makes it possible to extrapolate
our results to infinite volume.

For the results presented in this work we used the following equation to take finite
volume effects into account:
\begin{equation}
  \label{eq:finite_size}
  a^2 E =  \sum\limits_{n=0}^\infty E_n \beta^{-n} -
  \frac{1}{L^2}\sum\limits_{k=0} F_k \beta^{-k},
\end{equation}
where the coefficients $F_k$ are polynomials of order $k$ in $\ln (L)$. The details of the
infinite volume extrapolation are somewhat lengthy, and we postpone a more detailed discussion
of the extrapolation and its systematics to Appendix~\ref{sec:finite_V_effects}.

\begin{figure}[h]
  \centering
  \includegraphics[width=0.49\textwidth]{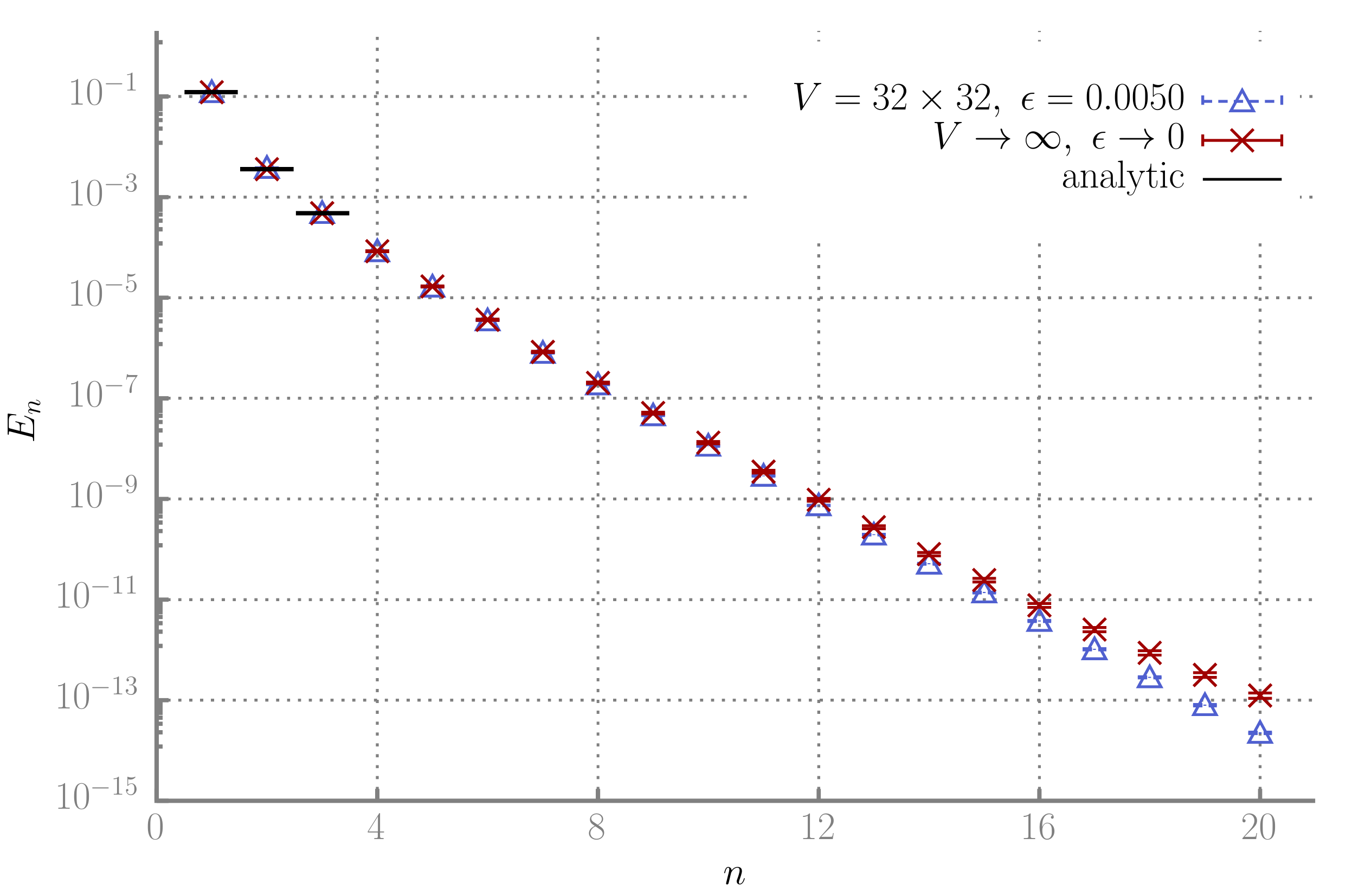}
  \caption{Expansion coefficients $E_n$ for the $PC(6)$ model from simulations at both,
    finite volume $V=32\times32$ and fixed discretization $\dtau=0.005$,
    as well as after the extrapolation to infinite volume and vanishing $\dtau$. The
    short horizontal bars mark the analytical weak coupling result. On the logarithmic
    scale, the error bars are smaller than the symbols.}
  \label{fig:N06_coeffs}
\end{figure}

Figure~\ref{fig:N06_coeffs} shows a synopsis of our numerical coefficients $E_n$ for
the $PC(6)$ model, i.e.,
a fixed
$N$. They perfectly meet the analytically known formulas for $E_{1,2,3}$ and have been
determined very precisely over many orders of magnitude. This figure also demonstrates the
importance of performing the extrapolation to $\dtau=0$ and to infinite volume: for high
orders the corrections to the expansion coefficients are extremely large and the finite
volume results are off by orders of magnitude.

It is interesting to note that the coefficients $E_n$ seem to fall off exponentially with
$n$. Up to the expansion orders we consider, the asymptotic nature of the expansion is
completely hidden by the large value of $n^*\approx 50$, only after which the coefficients
grow factorially; see Eq.~\eqref{eq:nstar}. With our current numerical setup, it is not
feasible to calculate expansion coefficients up to such high orders to directly observe
this growth of the $E_n$. This is at variance with four-dimensional gauge theories, where
$\beta_0=11$ and for the plaquette one computes a much smaller $n^*\approx 4/11$, such
that those coefficients grow from the start~\cite{Bali:2014fea}.

However, the derivation of $n^*$ assumes that the asymptotic behavior of
Eq.~\eqref{eq:En_asympt} of the expansion coefficients sets in before
the coefficients
start to grow; thus, the ratios $r_n$ of consecutive $E_n$ should approach the constant in
Eq.~\eqref{eq:rn_asympt} already for $n < n^*$. These ratios are therefore a more
sensitive signal for renormalons.
\begin{figure}[!t]
  \centering
  \includegraphics[width=0.95\linewidth]{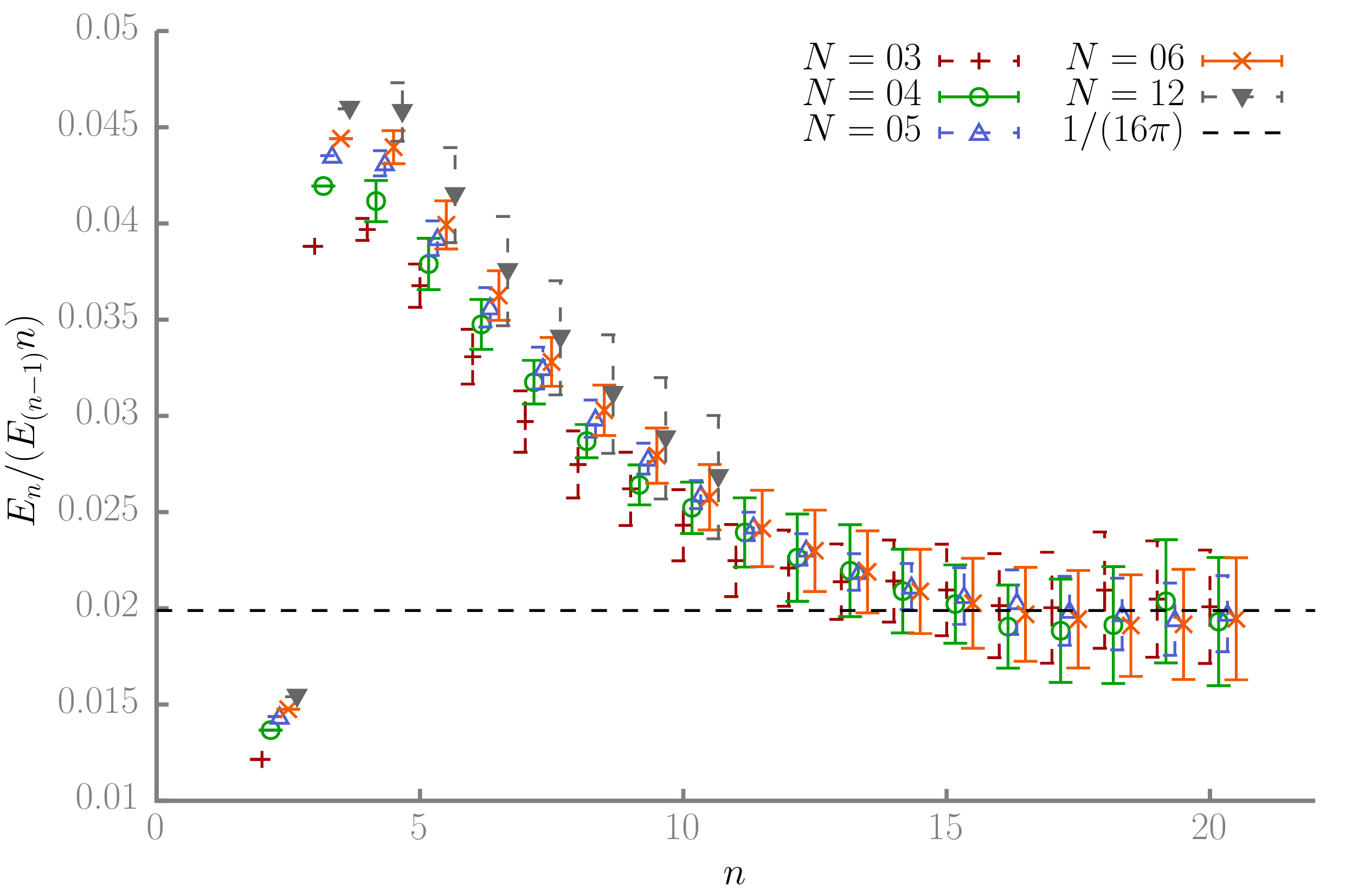}
  \caption{The ratios $r_n$ for various $N$ after the extrapolations
   $\dtau \to 0$ and $V \to \infty$. Results for rank $N > 3$ are slightly shifted
   horizontally for better visibility. The dashed line marks the (leading order)
   asymptotic $1/(16\pi)$, see Eq.~\eqref{eq:rn_asympt}.}
  \label{fig:ratios_Vinf}
\end{figure}

To calculate the ratios $r_n$, we first perform the extrapolation to $\dtau=0$ and then
use Eq.~\eqref{eq:finite_size} to take finite size effects into account. The results
are plotted in Fig.~\ref{fig:ratios_Vinf}.  We find that the asymptotic
behavior sets in somewhere around order $n=15$ independent of the rank $N$ in accordance
with the renormalon picture.

\begin{figure}[t]
  \centering
  \includegraphics[width=0.95\linewidth]{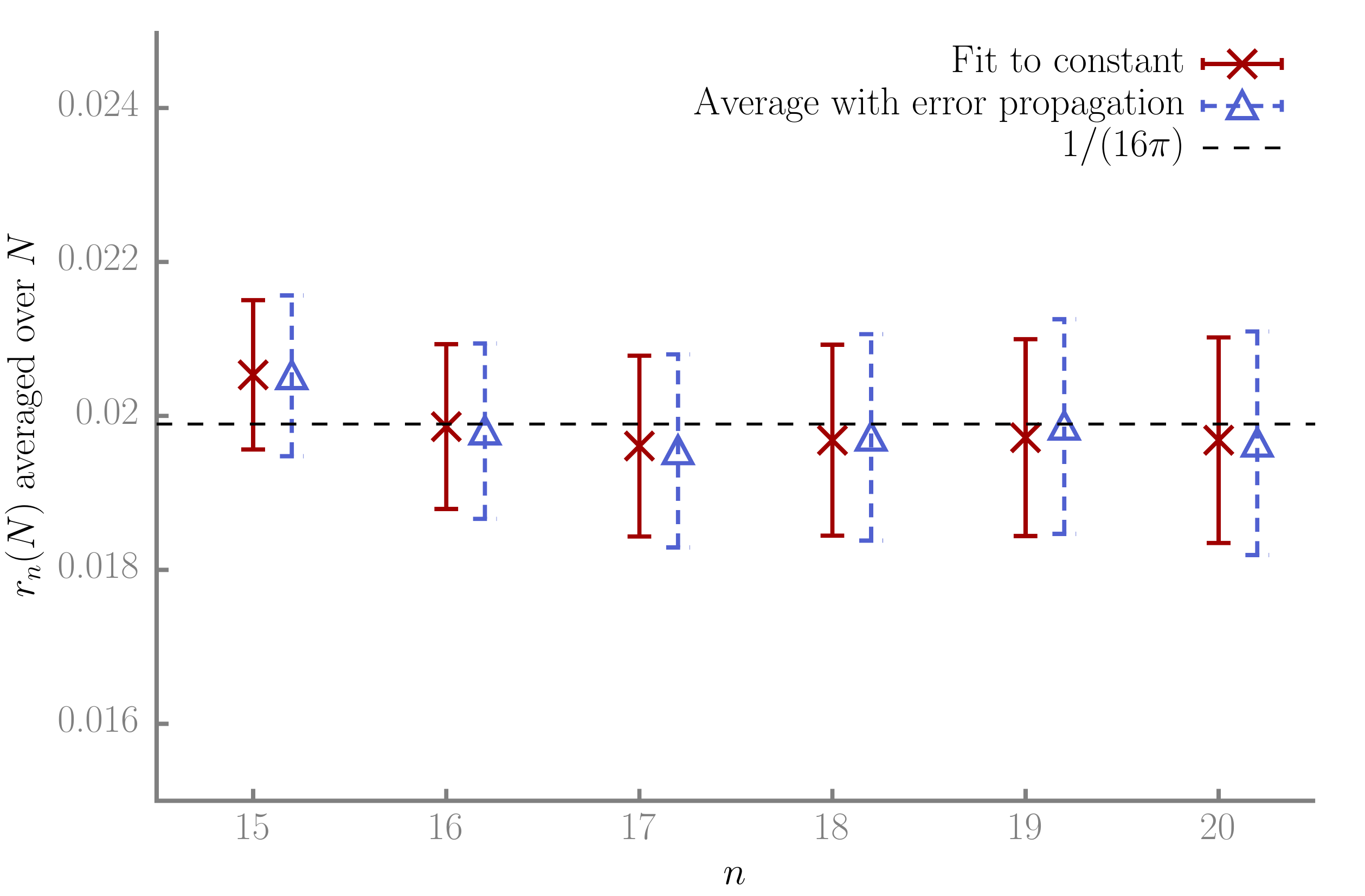}
   \caption{The plot shows the results of fitting  the ratios $r_n(N)$ over various $N$  to a constant for
     each $n$ in the region $n \geq 15$, where the asymptotic behavior has already set in.
     For comparison the results for averaging over $N$ with Gaussian error
     propagation, slightly shifted to the right for better visibility, are also shown.}
  \label{fig:ratios_Vinf_Nfit}
\end{figure}

While our data clearly show the expected asymptotic behavior, the error bars on the
ratios are relatively large. It is tempting to look for a plateau in the ratios $r_n$ to
perform a fit to a constant and get a better estimate for the asymptotic behavior of the
expansion coefficients. The problem with this approach is that subsequent ratios $r_n$ and
$r_{n+1}$ are strongly correlated, since the coefficient $E_n$ is used in the calculation
of both of them. The simulations and fits for different rank $N$ are, however,
independent. In Fig.~\ref{fig:ratios_Vinf_Nfit} we show the results of fitting a constant
to $r_n(N)$ for fixed $n$. The fit essentially averages over $N$ while also
taking the individual errors into account. The results for such a $N$-average
with Gaussian error propagation are shown for comparison.  For large expansion orders $n
\gtrsim 15$, after the asymptotic behavior sets in, the ratios $r_n$ should no longer
depend on $N$ and combining data for different ranks is justified. The plots in
Fig.~\ref{fig:ratios_Vinf_Nfit} are in very good agreement with the prediction from
Eq.~\eqref{eq:rn_asympt} with relative errors that are smaller than $10\%$.

To summarize, we have determined the perturbative coefficients of the energy in particular
two-dimensional sigma models by a suitable lattice technique -- implicitly summing up
factorially many diagrams -- including dedicated continuum and infinite volume
extrapolations. Using the first few analytically known coefficients as benchmarks, we have
determined up to 20 high-order coefficients which spread over many orders of
magnitude. Their ratios (divided by the order) clearly approach a constant consistent with
the renormalon picture. The latter is based mainly on the coupling dependence of the
strong scale (and the mass dimension of the quantity), thus reflecting nonperturbative
physics. Our \emph{ab initio} results validate the universality of the renormalon/resurgence
picture, i.e., the independence on the rank of the sigma model. The method utilized in
this study can easily be extended to twisted boundary conditions in one or even two
directions, which ought to reduce the volume dependence.  Likewise, asymptotic series in
other observables and/or other sigma models can be studied to shed more light on the
resurgence picture.

The NSPT code used in this work~\cite{Puhr1902}
as well as the original data and the data analysis scripts~\cite{Puhr1901} are
available on GitHub.

\bigskip
\noindent {\it Acknowledgments} -- We thank Gerald Dunne, Mithat \"Unsal and in particular
Gunnar Bali for useful discussions, and Marcos Mari\~no for pointing us to
reference~\cite{Volin:2009wr}. We are indebted to Jakob Simeth for his contributions to
the NSPT code used for our simulations. The computations were performed on a local cluster
at the University of Regensburg and on the Linux cluster of the LRZ in Garching. GNU
parallel~\cite{Tange:2018:01} was used to speed up the data analysis.  We acknowledge
support from DFG (Contracts No.\ BR 2872/6-2 and No.\ BR 2872/8-1).


\appendix

\section{NSPT for the Principal Chiral Model}
\label{sec:NSPT}
The framework of NSPT  applied in this work is not new, but not well-known outside the
lattice community. For the convenience of the reader, we therefore give a brief
introduction to NSPT in this section. A more thorough exposition can be found in the
referenced original works.
NSPT was first developed in the context of QCD in~\cite{DiRenzo:1994ym,DiRenzo:1994sy}
and a review can be found in~\cite{DiRenzo:2004hhl}.

\subsubsection{Numerical Perturbation Theory }
\label{subsec:NSPT}

The idea of numerical
perturbation theory is to formally perform a weak coupling expansion of the lattice fields
$U$ in powers of $\betapf{1}$ up to order $\beta^{-M}$,
\begin{equation}
  \label{eq:nspt_expansion}
 U = \id + U_1 \betapf{1} + U_2 \beta^{-1} + \cdots +
 U_{2M}\beta^{-M}, 
\end{equation}
where $\beta$ is the lattice coupling. In this work we only consider expansions around the
vacuum, where the leading term $U_0$ is given by the unit matrix on all lattice sites.

Algebraic operations with these truncated series are straightforward,
\begin{align}
  \label{eq:pert_sum}
  U + U' &= \sum\limits_{k=0}^{2M}(U_k + U'_k)\betapf{k}, \\
  \label{eq:pert_prod}
  U\cdot U' &= \sum\limits_{k=0}^{2M} \left(\sum\limits_{l=0}^k U_lU'_{l-k}\right) \betapf{k}. 
\end{align}
Once addition and multiplication are defined, any analytic function of the fields $U$ can
be evaluated by inserting the field expansion into the power series of the function.

The most expensive operation is the multiplication of fields, which requires
$\mathcal{O}(M^2)$ multiplications of coefficients~\footnote{This is only true for a naive
implementation of the convolution~\eqref{eq:pert_prod}. Using, e.g., a fast Fourier
transform algorithm for the convolution would require only $\mathcal{O}(M \log_2 M)$
multiplications.}. The numerical cost of an NSTP simulation is consequently roughly
proportional to $M^2$.  Compared to diagrammatic perturbation theory, where the number of
Feynman diagrams to be taken into account typically grows like $\mathcal{O}(M!)$, this is
very efficient.

\subsubsection{Stochastic Quantization on the Lattice}
\label{sec:stochquantlat}

An obvious precondition for the use of numerical perturbation theory is that all the
functions involved in the computations can be expanded in powers of $\betapf{1}$.  Most
state of the art lattice simulations use Monte Carlo methods based on the
Metropolis-Hastings algorithm to sample the configuration space. Metropolis-Hastings has a
big disadvantage from the point of view of numerical perturbation theory: it employs an
accept-reject step, where a proposed configuration update is accepted with a certain
probability depending on the change in the action. An accept-reject step cannot be
formulated as a function of powers of $\betapf{1}$ and is therefore not suitable for
numerical perturbation theory.

One alternative to Metropolis-Hastings type algorithms is stochastic
quantization~\cite{Damgaard:1987rr}, which is based on the Langevin equation. Stochastic
quantization introduces a new dimension, usually called the stochastic time $\tau$. The
evolution of the fields in $\tau$ is governed by the Langevin equation. For a scalar field
$\phi$ with action $\SA[\phi]$, the Langevin equation is given by
\begin{equation}
  \label{eq:scalar_langevin}
  \frac{\partial \scft}{\partial \tau} = - \frac{\delta \SA[\scft]}{\delta \scft} + \etab(x,\tau),
\end{equation}
where $\etab(x,\tau)$ is a Gaussian noise term with the properties 
\begin{align}
  \sbev{\etab(x,\tau)} &= 0 \label{eq:noise_mean} \\   
  \sbev{\etab(x,\tau)\etab(x',\tau')} &= 2\delta(x-x')\delta(\tau - \tau').   \label{eq:noise_var} 
\end{align}
In the equations above, $\sbev{\cdots}$ denotes the average over the noise $\etab$.

For numerical calculations, the partial differential equation~\eqref{eq:scalar_langevin}
has to be replaced by a finite-difference equation. Making $\tau$
discrete with a step size $d\tau=\dtau$ leads to
\begin{equation}
  \label{eq:discrete_langevin}
  \phi(x,\tau_{n+1}) = \phi(x,\tau_n) - F_x[\phi,\eta],
\end{equation}
where $-F_x[\phi,\eta]$ is a discretization of the $\tau-$integral of the right-hand side of
Eq.~\eqref{eq:scalar_langevin}. A choice with an error of order $\dtau$ is the Euler
method,~\footnote{It seems counterintuitive that the Euler method with a term
proportional to $\sqrt{\dtau}$ gives an error of order $\dtau$. The origin of the square
root term is the discretization of Eq.~\eqref{eq:noise_var}. Note that the dimension
of the Dirac $\delta$-function $[\delta(\tau-\tau')]=[\frac{1}{\tau}]=[\frac{1}{\dtau}]$,
whereas the Kronecker-delta in Eq.~\eqref{eq:var_discrete} is dimensionless. The
correct discretization of the  $\delta$-function is therefore $\delta(\tau-\tau') \to
\frac{1}{\dtau} \delta_{\tau_i,\tau_j}$. If we want to keep the variance of the
noise~\eqref{eq:var_discrete} independent of the step size $\epsilon$, we have to multiply
every occurrence of $\eta$ by $\dtau^{-1/2}$, i.e., $\etab \to \frac{\eta}{\sqrt{\dtau}}$.
In the Euler discretization this gives rise to the term $\dtau \frac{\eta}{\sqrt{\dtau}} =
  \sqrt{\dtau}\, \eta$.}
\begin{equation}
  \label{eq:Fx_Euler}
   F_x[\phi,\eta] := \dtau\, \frac{\delta \SA[\scft]}{\delta \phi(x,\tau_n)} - \sqrt{\dtau}\,\eta(x,\tau_n)
 \end{equation}
 with the Gaussian noise $\eta$, which has zero mean and obeys the discretized version of
 Eq.~\eqref{eq:noise_var}
 \begin{equation}
   \label{eq:var_discrete}
   \sev{\eta(x_i,\tau_n)\eta(x_j,\tau_m)} = 2\delta_{x_i,x_j}\delta_{\tau_n,\tau_m}.
 \end{equation}
It can be shown that for sufficiently large $\tau_n$ the field configurations
$\{\phi(x,\tau_n)\}$ produced by the discretized Langevin with $F_x$ given by
Eq.~\eqref{eq:Fx_Euler} are distributed with probability density
$\exp(-\bar{\SA}[\phi])$~\cite{Batrouni:1985jn}, where 
the equilibrium action $\bar{\SA}$
differs from the continuum action $\SA$ by terms of $\order{\dtau}$,
\begin{equation}
  \label{eq:equilibrium_action}
  \bar{\SA}[\phi] = \SA[\phi] + \dtau S_1[\phi] + \order{\dtau^2}.
\end{equation}
Evidently, discrete Langevin simulations suffer from a systematic error, since 
they sample
configurations with respect to a probability distribution that differs from the desired
one. This is a disadvantage of the Langevin ansatz and makes it necessary to
run simulations for several different values of $\dtau$ and to extrapolate the results to
$\dtau = 0$. On the other hand, the big advantage of stochastic quantization from the point
of view of numerical perturbation theory is the absence of an accept-reject
step. Stochastic quantization in combination with numerical perturbation theory goes by
the name of NSPT and can be used to numerically compute expansion coefficients on the
lattice.

For completeness, we mention a recently developed new formulation of NSPT, where the
Langevin equation is replaced by an stochastic molecular dynamics (SMD)
algorithm~\cite{DallaBrida:2017pex,DallaBrida:2017tru}. The SMD based NSPT has potential
advantages over the Langevin formulation, like smaller autocorrelation times. In
this work, we stick to the established Langevin method, which has proven itself in practice
in large scale NSPT simulations in the context of QCD.

\subsubsection{Discrete Langevin for Constrained Systems}
\label{sec:discrete_langevin}

So far, we have only discussed the simple case of an unconstrained scalar field $\phi$. The
generalization to fields which are elements of a Lie group is straightforward~\cite{Batrouni:1985jn, Catterall:1990qn}. For $U \in SU(N)$, the discretized
Langevin equation reads
\begin{equation}
  \label{eq:Lie_Langevin_discrete}
  U(x,\tau_{n+1}) = \exp\left(-i \sum\limits_a T^a F_x^a \right) U(x,\tau_n),
\end{equation}
where $T^a$ stands for the generators of $SU(N)$ (normalized to
$\tr(T^aT^b)=\delta_{a,b}/2$) and the generalization of the Euler term
is given by
\begin{equation}
  \label{eq:euler_group}
  F_x^a = \dtau \nabla_x^a \SA[U] + \sqrt{\dtau}\,\eta^a
\end{equation}
with the noise $\eta^a = \eta(a, x_i,\tau_i) T^a$ and the Lie derivative $\nabla_x^a$
defined via
\begin{equation}
  \label{eq:Lie_derivative}
  f(e^{i\sum\limits_a T^a\omega^a}U) = f(U) + \sum\limits_a \omega^a \nabla^a f(U) +
  \order{\omega^2}. 
\end{equation}
Note that $\nabla_x^a$ acts only on the field $U(x)$ at site $x$. The noise $\eta(a,
x_i,\tau_i)$ is Gaussian with zero mean and variance,
\begin{equation}
   \label{eq:var_discrete_group}
   \sev{\eta(a,x_i,\tau_i)\eta(b,x_j,\tau_j)} = 2\delta_{x_ix_j}\delta_{\tau_i\tau_j}\delta_{ab}.
 \end{equation}
 For the action~\eqref{eq:lat_action1} of the $PC(N)$ model, the Lie derivative is given by
   \begin{equation}
     \label{eq:Lie_deriv_PCN}
     \begin{split}
       \nabla_x^a \SA[U] &=  -i \beta N  \sum\limits_\mu\tr \Big( T^a U(x)\Ud(x+\mu)  \\
       & \qquad \qquad -  U(x+\mu)\Ud(x)T^a \\
       & \qquad \qquad - U(x-\mu)\Ud(x)T^a \\
       & \qquad \qquad  + T^a U(x)\Ud(x-\mu)\Big)
     \end{split}
   \end{equation}
Exploiting the property
 \begin{equation}
   \label{eq:SUN_prop}
    T_{ij}^aT^b_{kl} = \frac{1}{2}\left(\delta_{il}\delta_{jk} - \frac{1}{N}\delta_{ij}\delta_{kl}\right)
  \end{equation}
of the $SU(N)$ generators, the sum over Lie
derivatives in the exponent of Eq.~\eqref{eq:Lie_Langevin_discrete} can be performed
analytically,
  \begin{multline}
    \label{eq:Lie_derivative_sum}
    \sum\limits_a \nabla_x^a \SA[U] T^a = -i\frac{\beta N}{2}\sum\limits_\mu 
    \Big(\Mmu(x) - \Mmud(x) \\  - \Mmu(x-\mu) + \Mmud(x-\mu)\Big), 
  \end{multline}
where $\Mmu(x) := U(x)\Ud(x+\mu) - \frac{1}{N}\tr\big(U(x)\Ud(x+\mu)\big)$. Note that
$\Mmu(x) - \Mmu^\dagger(x)$ is an anti-Hermitian traceless matrix and consequently $\exp\big(i \sum_a
\nabla_x^a \SA[U] T^a \big) \in SU(N)$, as it should be to ensure that the updated field
is also an element of the group.

The use of the Euler integrator \eqref{eq:euler_group} again leads to systematic errors of
order $\order{\dtau}$. Runge-Kutta type algorithms for discrete Langevin updates of $SU(N)
$-valued fields have been discussed in the
literature~\cite{Batrouni:1985jn,Torrero:2008vi,Bali:2013pla}. In this
work we employ the algorithm from~\cite{Bali:2013pla}, which is of
order $\order{\dtau^2}$. We find that the higher numerical cost in comparison to the Euler
method is offset by the possibility to use larger values of $\dtau$ in the simulations.

Having defined a discrete Langevin equation for fields $U$, it seems straightforward to
apply numerical perturbation theory to this framework. There is, however, a subtle
issue. Plugging the expansion~\eqref{eq:nspt_expansion} into
Eq.~\eqref{eq:Lie_deriv_PCN}, it becomes clear that the force starts with terms of
order $\beta^{\frac{1}{2}}$, whereas the noise term in Eq.~\eqref{eq:Fx_Euler} is of
order $\order{1}$ and the fields are expansions in
$\betapf{1}$~\footnote{We use the Euler method as an
example here, but the situation is the same for Runge-Kutta algorithms.}.  Evidently, this
leads to inconsistencies. The solution is to redefine the time step $\dtau \to
\dtau':=\beta\dtau$. By absorbing $\beta$ in the time step, the force term starts with
order $\betapf{1}$. Note that the noise is now also of order $\betapf{1}$ and that the
leading-order term of the fields is not changed during a Langevin update. Since the
leading-order term is the point around which the fields are expanded --- in our case the
vacuum --- this is what we should expect.

\section{Simulation Details}
\label{sec:simdetails}
\subsubsection{Lattice Setups }
\label{subsec:latsetup}
For our simulations, we use only symmetric lattices with $V=L\times L$ and a lattice
spacing $a$ that is the same in all directions. All simulations are performed with at
least three different values of $\dtau$ for every lattice geometry considered; see
Table~\ref{tab:geometries} for details about the lattice setups.
As discussed in the previous section, the noise enters the NSPT Langevin equation only at
order $\betapf{1}$. It takes several Langevin sweeps before the noise can affect
the higher order terms of the field expansion. Moreover, from Eqs.~\eqref{eq:pert_sum}
and~\eqref{eq:pert_prod}, it is obvious that higher order coefficients have no influence on
lower order terms in NSPT calculations. As long as the field coefficients of a given order
$l$ are not thermalized, it makes therefore no sense to update terms of order $m > l$. The
initial condition of our simulations corresponds to a ``cold start'', where all
fields are equal to $\id$. 
To initialize the higher orders, we start with an expansion up to order two (in
$\betapf{1}$) and perform $\ninit=500$ Langevin sweeps. Then we increase the expansion
order by one and run the simulation for another $\ninit$ time steps. This is repeated
until we have reached the desired expansion order $M$, after which we perform a final
$\ninit + 1/\dtau$ initialization sweeps.

\begin{table}[bth]
   \begin{tabular}{llc}
     \toprule
     Order $M$ & Rank $N$ & Lattice extensions $L$ $(V = L \times L$) \\
     $20 (\beta^{-10})$ & $12$ &  $8, 12, 16, 20, 24, 32, \mathbf{48}  $\\
     $40 (\beta^{-20})$ & $3,4,5,6$ & $8, 12, 16, 20, 24, 32, \mathbf{48} $\\
     \botrule
  \end{tabular}
  \caption{Lattice setups for the Langevin runs. The simulations use five
    different stochastic time step sizes \mbox{$\dtau=0.01, 0.00875, 0.0075, 0.00625,
      0.005$}, except for the geometries printed in bold face, where $\dtau=0.01,0.0075, 0.005$.} 
  \label{tab:geometries}
\end{table}

We emphasize that our lattice fields only fulfill the constraint $U \in SU(N)$ up to the
current expansion order. When we go to a higher order, the unitarization is no longer
valid and we have to unitarize the fields up to the new order. It is not straightforward
to enforce $U \in SU(N)$, because it is in general not easy to see how the constraints
$\det(U)=1$ and $U^\dagger U = \id$ 
translate to conditions for the expansion coefficients
$U_k$. A well-known trick is to work with fields $A \in \mathfrak{su}(N)$ from the algebra
instead. Any element of $SU(N)$ can be written as
\begin{equation}
  \label{eq:SUN_exp}
  U = e^{i A} 
\end{equation}
for a suitable $A$. For $U \in SU(N)$ the field $A = \sum A_k \betapf{k} $ has to 
fulfill $A=A^\dagger$ and $\tr(A)=0$, which can be achieved by making every $A_k$ individually
Hermitian and traceless,
\begin{equation}
  \label{eq:A_normalise}
  A_k \to \frac{1}{2}\left(A_k + A_k^\dagger \right), \qquad A_k \to A_k - 
  \frac{\tr
  (A_k)}{N}
  \cdot \id_n\,.
\end{equation}
To go back and forth between the fields  $U$ and $A$ one utilizes the series expansion of
$\ln(1+x)$ and $\exp(x)$, respectively.  Taking
advantage of the expansion around the vacuum, which fixes $U_0 = \id$ and $A_0 = 0$,
considerably simplifies the calculations.
By taking the route over the auxiliary fields $A$, normalizing $U$ becomes
straightforward, albeit numerically expensive.

Once the lattice is initialized, we perform $\nup=500/\dtau$ Langevin sweeps with
measurements every $\nmeas = 100$ sweeps.  Note that we chose the same stochastic time
duration for all our simulations, since we expect the autocorrelation to depend on the
stochastic time, not the number of discrete Langevin time steps.

\subsubsection{Constraint Violation in Langevin Runs}
\label{sec:constraint}
In practical calculations, it is important to keep in mind that the NSPT algorithm described
in the last section respects the constraint $U \in SU(N)$ only if expressed in infinite
precision arithmetic. Round-off and numerical errors can lead to a violation of unitarity.

During our runs, we periodically check if the fields still fulfill the
constraints. Calculating $A$ every time would be very expensive and considerably 
slow down the simulations. We therefore check, order by order in the expansion, if the
fields $U$  are still unitary. To this end, we calculate
\begin{equation}
  \label{eq:Constraint_dev}
    \Delta_k = 
    \begin{cases}
      |\sqrt{N} - \|V_k\|_F |/N  & k = 0 \\
      \|V_k\|_F/N & k > 0
    \end{cases}, 
\end{equation}
with $V_k$ the expansion coefficients of unitarity,
\begin{equation}
  \label{eq:Constraint_dev_V}
  \sum_k V_k \betapf{k} := U U^\dagger = \sum\limits_{k=0} \left(\sum\limits_{l=0}^k
      U_lU^\dagger_{l-k}\right) \betapf{k}\,,
  \end{equation}
where $\| \cdot \|_F$ stands for the Frobenius matrix norm. The expectation value
$\vev{\Delta_k}$ (over all lattice sites) is taken as an estimate of the violation of the
constraint $U \in SU(N)$.

\begin{figure}[htb]
  \centering
  \includegraphics[width=0.49\textwidth]{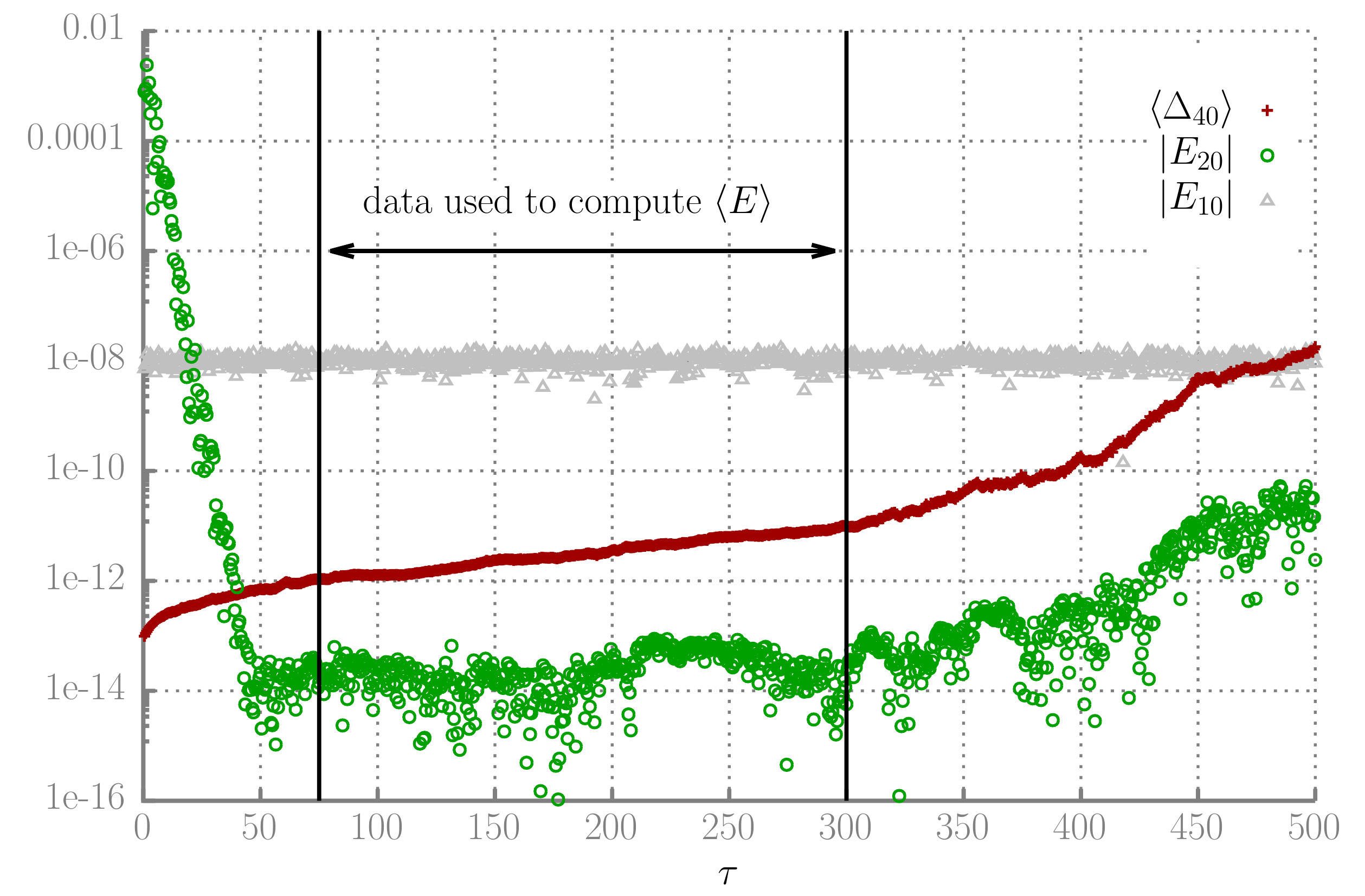}%
  \caption{Stochastic time history of the coefficient $E_{20}$ and the expectation value
    of the corresponding unitarity violation estimate $\vev{\Delta_{40}}$ for the
    parameters $N=6 , \dtau=0.005$ and $V=20\times20$. The thermalization phase up to $\tau
    \sim 50$ is clearly visible. For late $\tau \gtrsim 350$, the unitarity violation becomes
    large and seems to affect the equilibrium distribution of $E_{20}$. $E_{10}$, which is
    already thermalized after the initialization, is plotted for comparison.
    Our statistics only use the data marked in the plot.}  
  \label{fig:TimeSeries_N06}
\end{figure}

Monitoring the Langevin histories shows that the initialization is in general
not sufficient to thermalize the highest order coefficients of the energy density. We do
not observe any significant violation of unitarity for the lattices with $V \geq
32\times32$, but the small lattices with $V \lesssim 20\times20$ show large
$\vev{\Delta_k}$ even for moderate Langevin times $\tau$. The deviation from unitarity
strongly depends on the lattice volume and on the rank $N$. In general, all other
parameters being equal, $\vev{\Delta_k}$ is larger for higher rank $N$ and higher expansion
order $k$.

\begin{table}[tbh]
  \begin{tabular}[c]{cll}
    \toprule
    $V$ & $\tau$ range & Remarks \\
    \colrule 
    $16\times16$ & $100 \dash 500 $ & Only $N=3$ and w/o $\dtau=0.01, 0.0625$ \\
    $20\times20$ & $\;\; 75 \dash 300 $ & $N=6$  w/o  $\dtau=0.0075$  \\
    $24\times24$ & $150 \dash 400$ &  \\
    $
    \begin{matrix}
      32\times32 \\
      48\times48
    \end{matrix}
$  & $\left.
     \begin{matrix}
       170 \dash 500 \\
       220 \dash 500
     \end{matrix}\right\} 
$  &  No unitarity violation found \\
    \botrule 
  \end{tabular}
  \caption{Range of $\tau$ where the simulations are thermalized and unitarity violations are
    small. }
  \label{tab:data_errorfree}
\end{table}

For the final computation of the expectation values $\vev{E_n}$, we make sure to choose
only data from a stochastic time interval where the simulation is equilibrated and
unitarity violations are small. An example for $N=6$ is shown in
Fig.~\ref{fig:TimeSeries_N06}. Remember that the $E_n$ are coefficients for an expansion
in $\beta^{-1}$, whereas the index on the $\Delta_k$ corresponds to an expansion in
$\betapf{1}$.

To be consistent, for a given lattice size, we choose the
same $\tau$ interval for all orders and all parameter sets. In
Table~\ref{tab:data_errorfree} we summarize which data we used in the final fits. For
convenience, we simply call this data set \emph{setup~I} from now on.

The choice of which data to include is somewhat arbitrary and one should verify that it
does not influence the final results. As a cross-check, we consider two additional
datasets. The first consists of all the setups in Table~\ref{tab:geometries} and for the
second one we use the configurations of Table~\ref{tab:data_errorfree}. In both cases, we
take data from the stochastic time interval $\tau \in [250,500]$. For later use, we will
refer to these as \emph{setup~II} and \emph{setup~III}, respectively.  Remarkably, as will
be discussed in more detail in the next section, within uncertainties the results for the
ratios $r_n$ for all three datasets overlap. The reason is probably that for large
lattices with $V \gtrsim 24\times 24$, which seem to have a higher weight in the infinite
volume extrapolation, unitarity violations are very small.

\subsubsection{Extrapolation $\dtau \to 0$}
\label{sec:epstozero}
\begin{figure}[htb]
  \centering
  \includegraphics[width=0.245\textwidth]{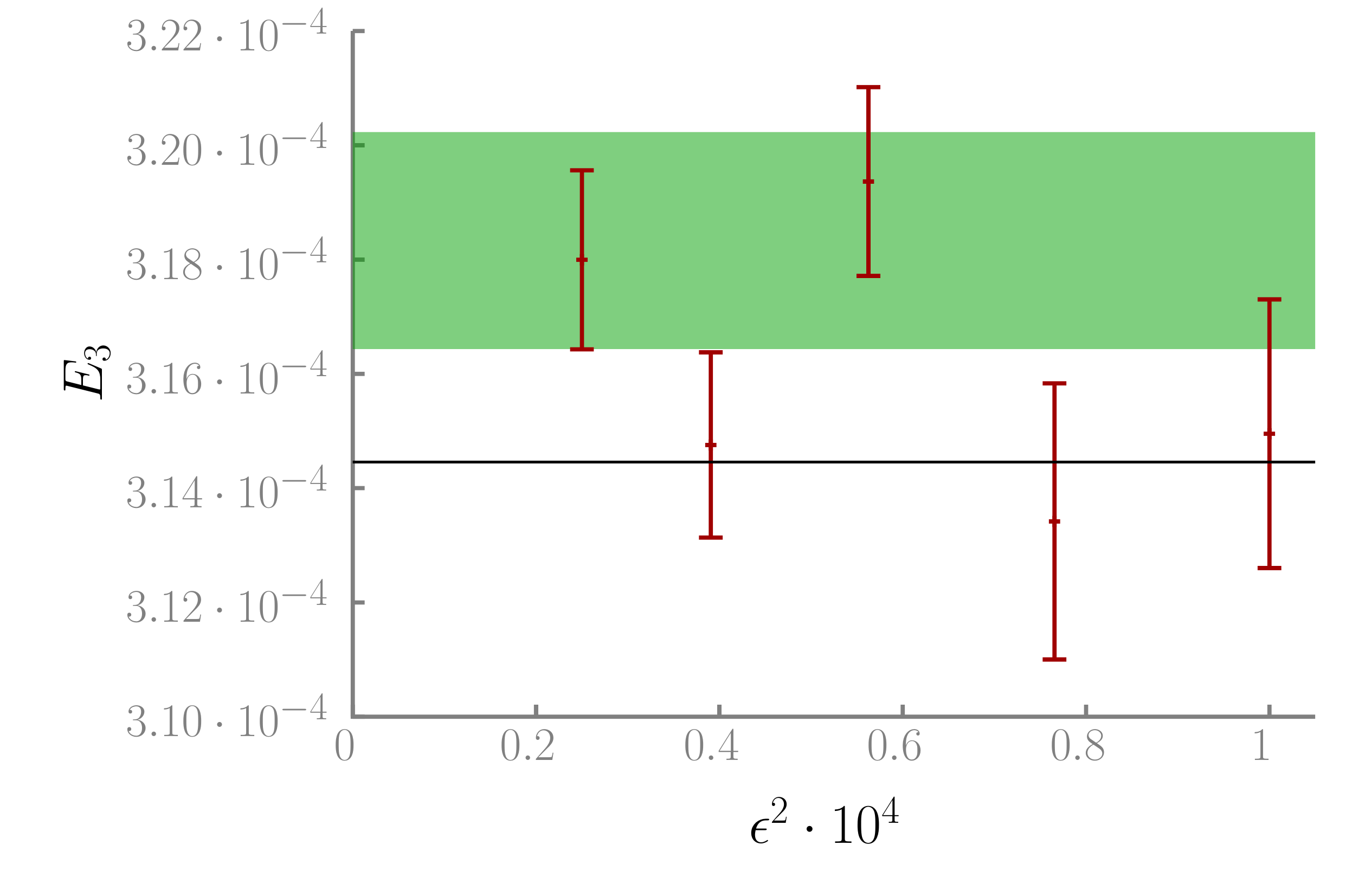}%
  \includegraphics[width=0.245\textwidth]{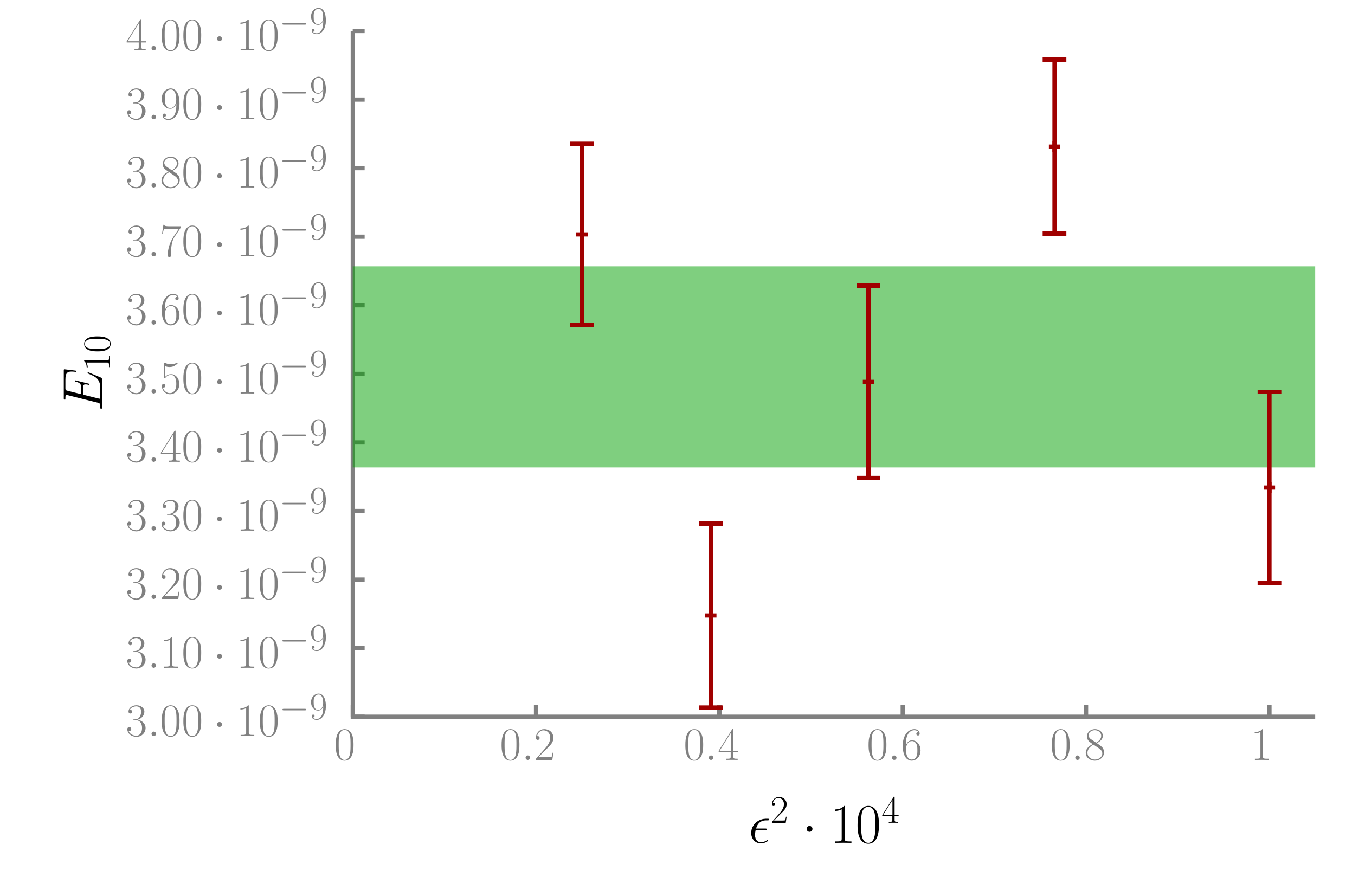}\\%
  \includegraphics[width=0.245\textwidth]{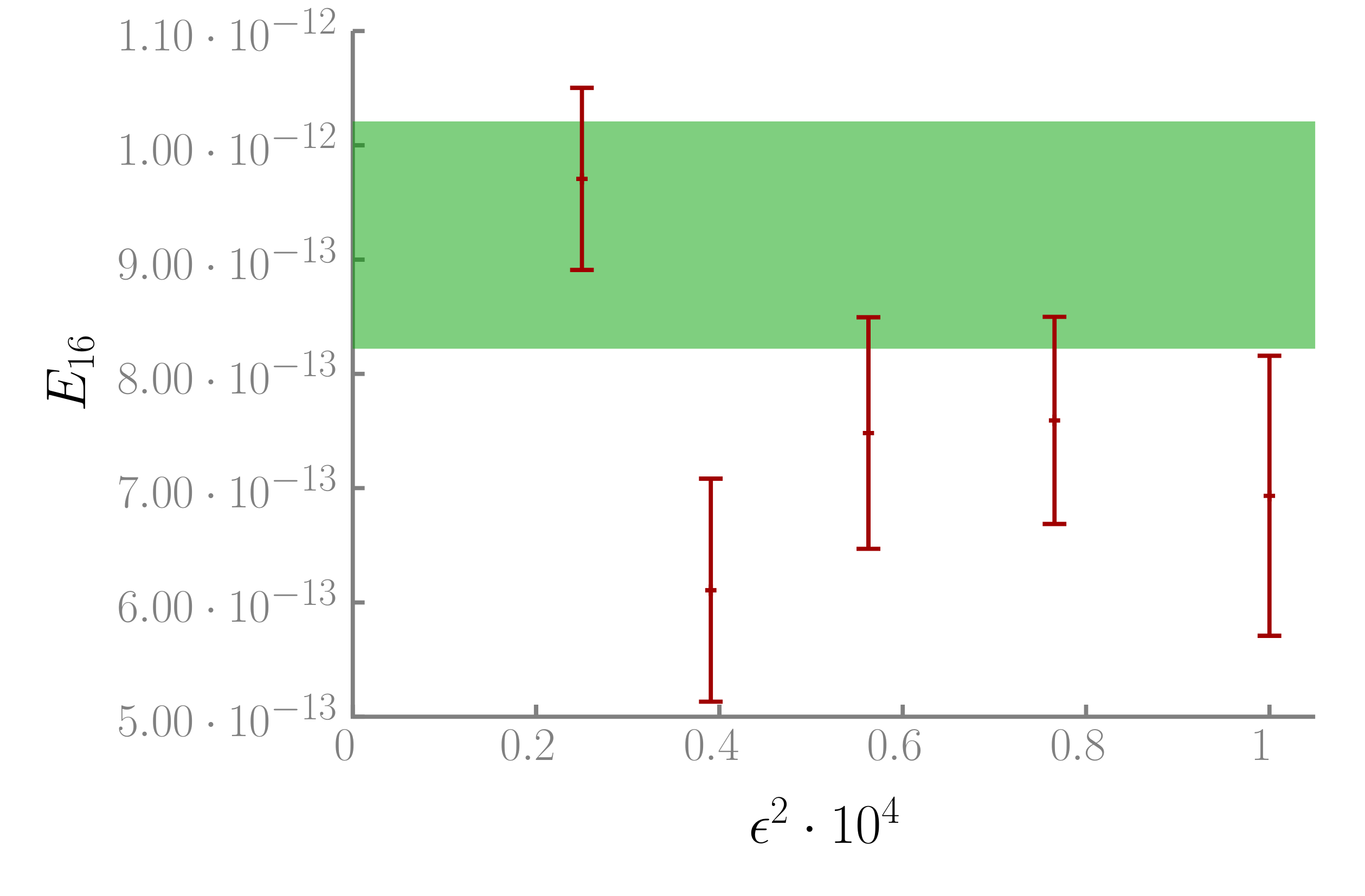}%
  \includegraphics[width=0.245\textwidth]{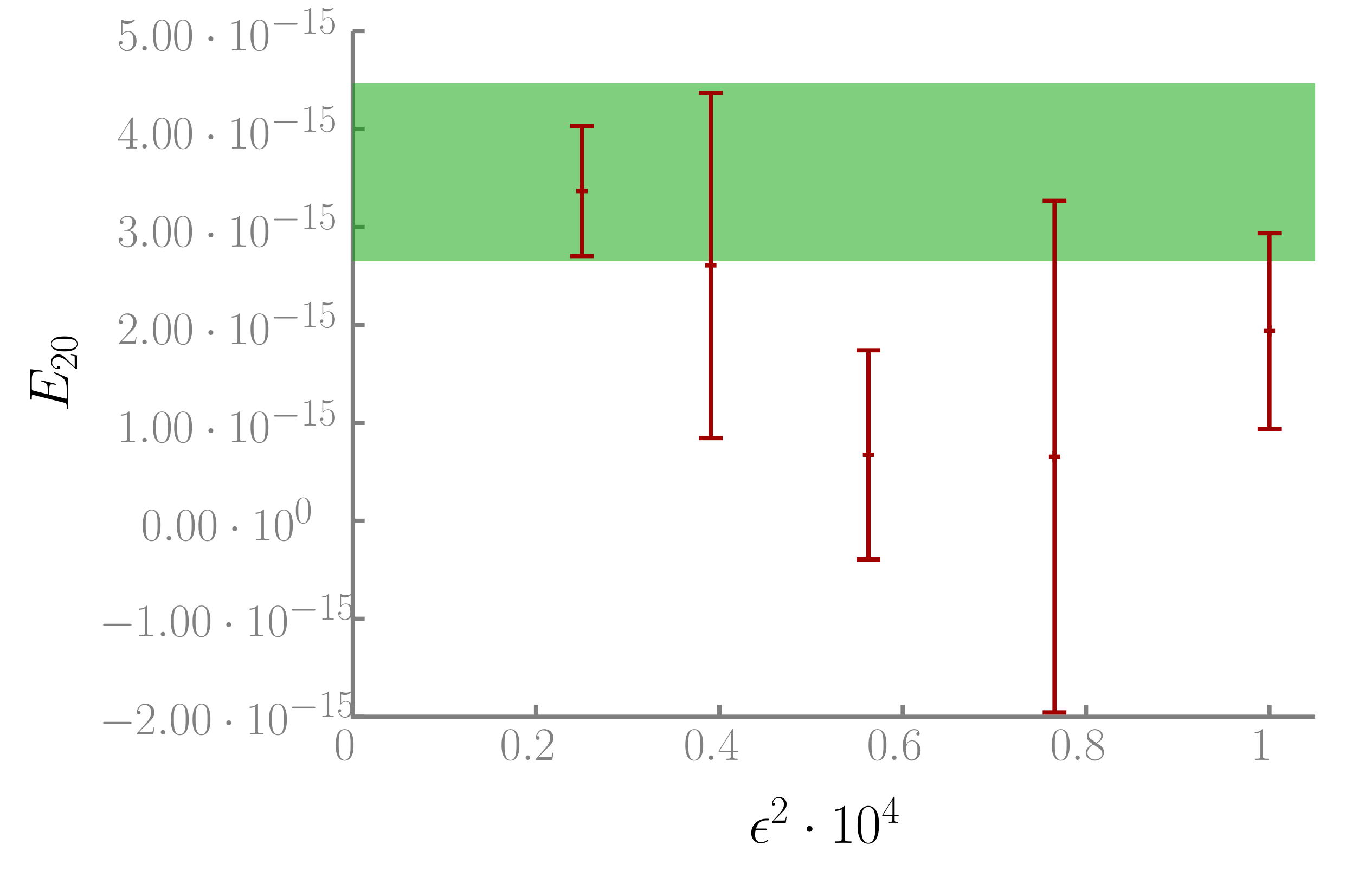}%
  \caption{Example for the extrapolation to $\dtau=0$. The plots show results from runs
    with $N=3$ and $L=32$ for $V=32 \times 32$. The expansion coefficients $E_n$ for the orders $n=3,10,16,20$
    are shown as a function of $\dtau^2$. The shaded region
    shows the result with uncertainties for an extrapolation for $\dtau \to 0$ obtained
    by a linear fit in $\dtau^2$. For $n=3$, the known perturbation theory result is
    plotted as a straight black line. The fit uncertainties do not include systematics
    from neglecting higher orders.}
  \label{fig:extrap_eps_zero}
\end{figure}

The Runge-Kutta integrator employed in the
simulations is of second order. The extrapolation to $\dtau=0$ can therefore be performed
by fitting a function linear in $\dtau^2$ to our data.
The systematics of neglecting higher order terms in the fit function are estimated to be
\begin{equation}
  \label{eq:syst_error} s_n = \left| 1 - \frac{E_n(\dtau \to 0)}{E_n(\dtau^\star)}\right|,
\end{equation}
where $\dtau^\star$ is the largest $\dtau$ value used for the given lattice setup which is
not larger than the median of the $\dtau$ values available.
Results for a run with $N=3$ and $L=32$ are shown in Fig.~\ref{fig:extrap_eps_zero}.
In our estimates for the statistical uncertainties of the expansion coefficients we take
the autocorrelation into account. The integrated autocorrelation time is estimated using
the automatic windowing algorithm from~\cite[Appendix C]{Madras:1988ei} with $c=5$.
Where available, we compare the NSPT coefficients with results from analytic perturbation theory
and find good agreement. The final uncertainty for the expansion coefficients extrapolated to
$\dtau=0$ is obtained by adding the uncertainties from the fit and the systematics in
quadrature.

\section{Finite volume effects}
\label{sec:finite_V_effects}
The extrapolation to infinite volume is based on OPE methods and we closely follow
reference~\cite{Bali:2014fea}. The energy density is symmetric under an exchange of $a
\leftrightarrow -a$ and the same is also true for our lattice action. Additionally, we
only consider symmetric lattices with equal extend in all directions. It follows that
the finite volume effects can only depend on even powers of $L$ and can be parametrized as
\begin{equation}
  \label{eq:finite_V_effects}
  E_n (L) = E_n^\infty - \frac{F_n(L)}{L^2} + \order{\frac{1}{L^4}},
\end{equation}
where $E_n^\infty$ is the infinite volume result.

\subsubsection{OPE of the Energy Density}
\label{sec:opefinitevol}
The lattice regularization introduces two distinct scales: the
inverse lattice spacing $1/a$ and the inverse of the linear lattice extent $1/(La)$. In
the infinite volume limit $ aL \gg a$, the two scales are well separated and it makes sense
to apply an OPE. The Wilson coefficients then depend on the hard modes of scale $1/a$, whereas
the soft modes of scale $1/(La)$ can be described in terms of expectation values of local
operators. Formally, we can then expand the expectation value of the perturbative energy
density as
\begin{equation}
  \label{eq:OPE_pert}
  \vev{E}_{\text{pert}} = E_{\text{pert}}(\alpha) + a^2 C_2(\alpha) \vev{O_2}_\text{soft}
  + \order{\frac{1}{L^4}},
\end{equation}
where
\begin{equation}
  \label{eq:E_pert}
  E_{\text{pert}}(\alpha) = \sum\limits_{n=0}^{\infty} E_n^{\infty} \alpha^n,
\end{equation}
with the infinite volume expansion coefficients $E_n^{\infty}$. The expectation value
$\vev{O_2}$ has to be proportional to $1/(La)^2$  on dimensional grounds and we write it as
\begin{equation}
  \label{eq:O_2_exp}
  a^2 \vev{O_2} := -\frac{1}{L^2} \sum \limits_{k=0} f_k \alpha^k(1/(La)). 
\end{equation}
Finally, absorbing a constant factor into the definition of the $f_k$, the Wilson
coefficient is 
\begin{equation}
  \label{eq:wilson_C_2}
  C_2(\alpha) := 1 + \sum\limits_{i>0} c_i \alpha^i(1/a). 
\end{equation}
With the help of the equations above, we can write $\vev{E_{\text{pert}}}(L)$ in two
different ways,
\begin{equation}
  \label{eq:FV_expansion}
 \vev{E_{\text{pert}}}(L) =  \sum\limits_n\left(E_n^{\infty} - \frac{F_n(L)}{L^2} \right)\alpha^n(1/a)
\end{equation}
and
\begin{equation}
  \label{eq:FV_expanison_2}
  \begin{split}
    \vev{E_{\text{pert}}}(L) &= \sum\limits_n E_n^{\infty} \alpha(1/a)^n \\
    & \quad  -\frac{1}{L^2}\left( 1 + \sum\limits_{i>0} c_i \alpha^i(1/a) \right) \\
    & \qquad \times \sum \limits_{k=0} f_k \alpha^k(1/(La)) + \order{\frac{1}{L^4}}.
  \end{split}
\end{equation}
Note that we are using a different index convention than~\cite{Bali:2014fea} here.

We are interested in calculating the coefficients $E_n^{\infty}$ from the finite volume
results $E_n(L)$. To this end, we can use the $\beta$-function of the $PC(N)$
model~\cite{Rossi:1993zc,Gonzalez-Arroyo:2018aus},
\begin{equation}
  \label{eq:beta_PC(N)} \beta(\alpha(\mu)) = \frac{d\alpha(\mu)}{d \ln(\mu)} = -
\alpha^2(\mu)\left(\beta_0 + \beta_1 \alpha(\mu) + \cdots \right)
\end{equation} 
to expand $\alpha^k(1/(La))$ around $\alpha^k(1/a)$ in
Eq.~\eqref{eq:FV_expansion}. Comparing \eqref{eq:FV_expansion} and
\eqref{eq:FV_expanison_2} order by order in $\alpha^k(1/a)$ finally yields expressions for
the functions $F_n(L)$ in terms of the coefficients $f_k, c_i$ and the $\beta$-function.
In any volume, the leading order of the expansion --- the expansion point ---  is fixed and
therefore $f_0=F_0(L)=0$. The first few nontrivial functions read:
\begin{equation}
  \label{eq:F_1}
  F_1(L) = f_1,
\end{equation}
\begin{equation}
  \label{eq:F_2}
    F_2(L) = (f_2 + c_1f_1) + f_1\beta_0 \ln(L), 
\end{equation}
and 
\begin{equation}
    \label{eq:F_3}
    \begin{split}
      F_3(L) &= (c_1f_2 + c_2f_1 + f_3) \\
      & \quad + \Big( \beta_0(c_1f_1 + 2f_2) + \beta_1f_1\Big)\ln(L) \\
      & \quad + f_1 \beta_0^2\ln^2(L).  
    \end{split}
\end{equation}
In general, $F_k(L)$ is a polynomial of degree $(k-1)$ in $\ln(L)$ with coefficients
depending on $\{f_l\}_{l \leq k}$, $\{c_j\}_{j \leq (k-1)}$, and on the $\beta$-function
(via the coefficients $\beta_0, \beta_1, \cdots$).

A simple closed expression to generate $F_k(L)$ for given $k$ is not known. We use
SymPy~\cite{Meurer:2017yhf} to explicitly compute the expansion of
Eq.~\eqref{eq:FV_expanison_2} up to order $\alpha^{20}$, from which the $F_k(L)$ can
be read off.

\subsubsection{Fitting the Volume Dependence}
\label{sec:fits}
Knowing the functional form of the finite volume effects, we can perform fits to our
data to extract the infinite volume coefficients. In its most generic form, the fit
function we use is given by
\begin{equation}
    \label{eq:fit_generic}
    E_n(L) = E_n^\infty - \frac{F_n(L)}{L^2}, 
\end{equation}
where $F_n$ depends on the unknown fit parameters $\{f_k\}$ and $\{c_i\}$.
  
The $f_k$ and $c_i$ couple finite size effects for different expansion orders. We follow the
ansatz of~\cite{Bali:2014fea} and perform a simultaneous fit to all expansion coefficients. If
the fields in the NSPT are expanded up to order $M$, the coefficients $c_{M-1}$ and $f_M$
solely appear in $F_M$ and only as constant terms in the polynomial in $\ln(L)$. Therefore,
from the perspective of the fit, one of them is redundant and we set $c_{M-1}=0$. The
coefficients up to $E_3^\infty$ are known from perturbation theory and are used as input
values in the fits. All in all, we end up with a fit function with $3M-5$ free parameters:
$(M-3)$ from the infinite volume coefficients $E_n^\infty$, $M$ parameters $f_k$, and
$(M-2)$ unknown $c_i$.
 
In all our fits we neglect higher order terms in the $\beta$-function and set $\beta_{i>1}
= 0$. The leading coefficients are set to their known (and regularization independent)
values. The systematic error of truncating the $\beta$-function is estimated by performing
separate fits where $\beta_1$ is also set to zero.
In our fits we observe a behavior that has also been noticed in~\cite{Bali:2014fea}: in the
functions $F_n$, the terms containing $c_if_{j-i}$ for fixed $j$ and different $i$ are hard to
distinguish in the fitting procedure. The reason can be understood by looking at
Eq.~\eqref{eq:FV_expanison_2}: the running of the terms
$c_if_{j-i}\alpha(1/(La))^i\alpha(1/a)^{j-i}$ is very similar for fixed $j$, especially if
$j$ is small. This introduces strong correlations between such terms in the fit and
ultimately leads to large uncertainties in the infinite volume expansion coefficients.
Moreover, when we include systematic errors for the extrapolation $\dtau \to 0$ in the
uncertainties for $E_n(L)$, the fits yield unrealistically small $\chi^2$ values. Our
estimate for the systematic errors is probably too conservative and does not put strong
constraints on the fit parameters.  As a cross-check, we repeat the fits with only the
statistical uncertainties included.

The vastly different scales of the $E_n$ for different expansion orders make it
hard to compare the coefficients directly. Instead, we consider the ratios $r_n$ obtained
with the infinite volume coefficients $E_n^\infty$ from the fits. Uncertainties
for the ratios are calculated with Gaussian error propagation from the uncertainties in
$E_n^\infty$.

\begin{figure}
  \centering
  \includegraphics[width=0.40\textwidth]{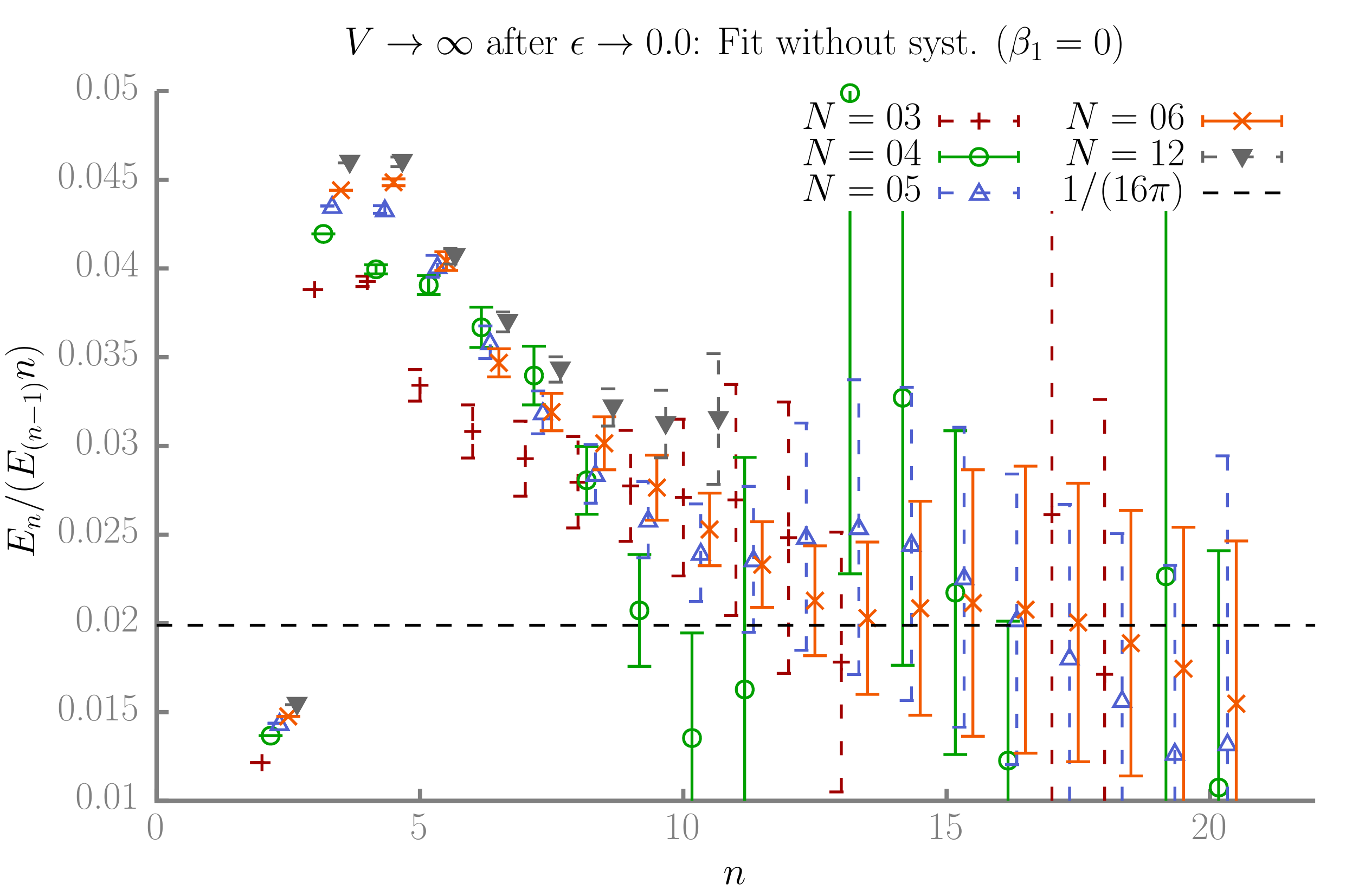}%
  \includegraphics[width=0.40\textwidth]{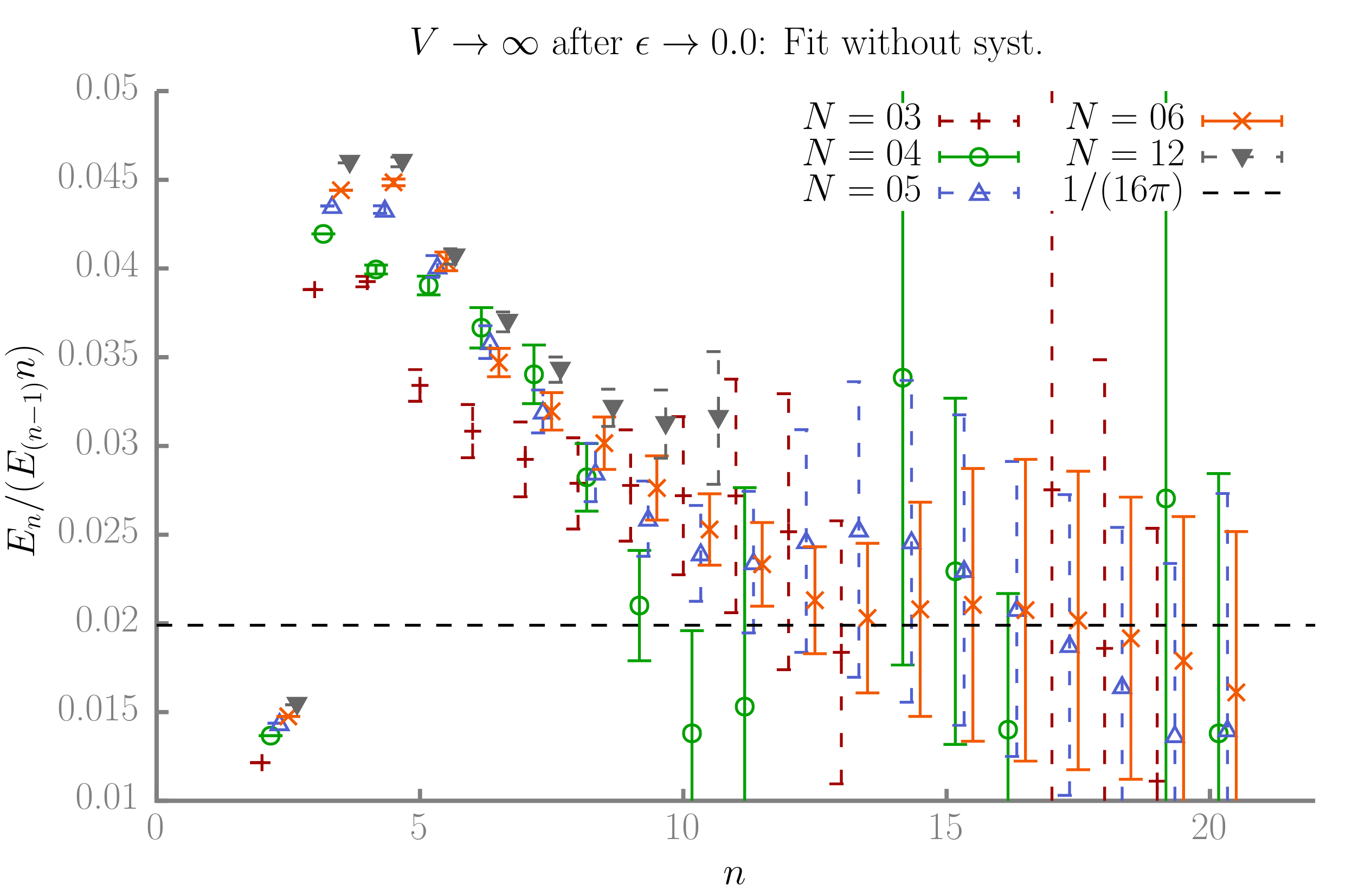} \\%
  \includegraphics[width=0.40\textwidth]{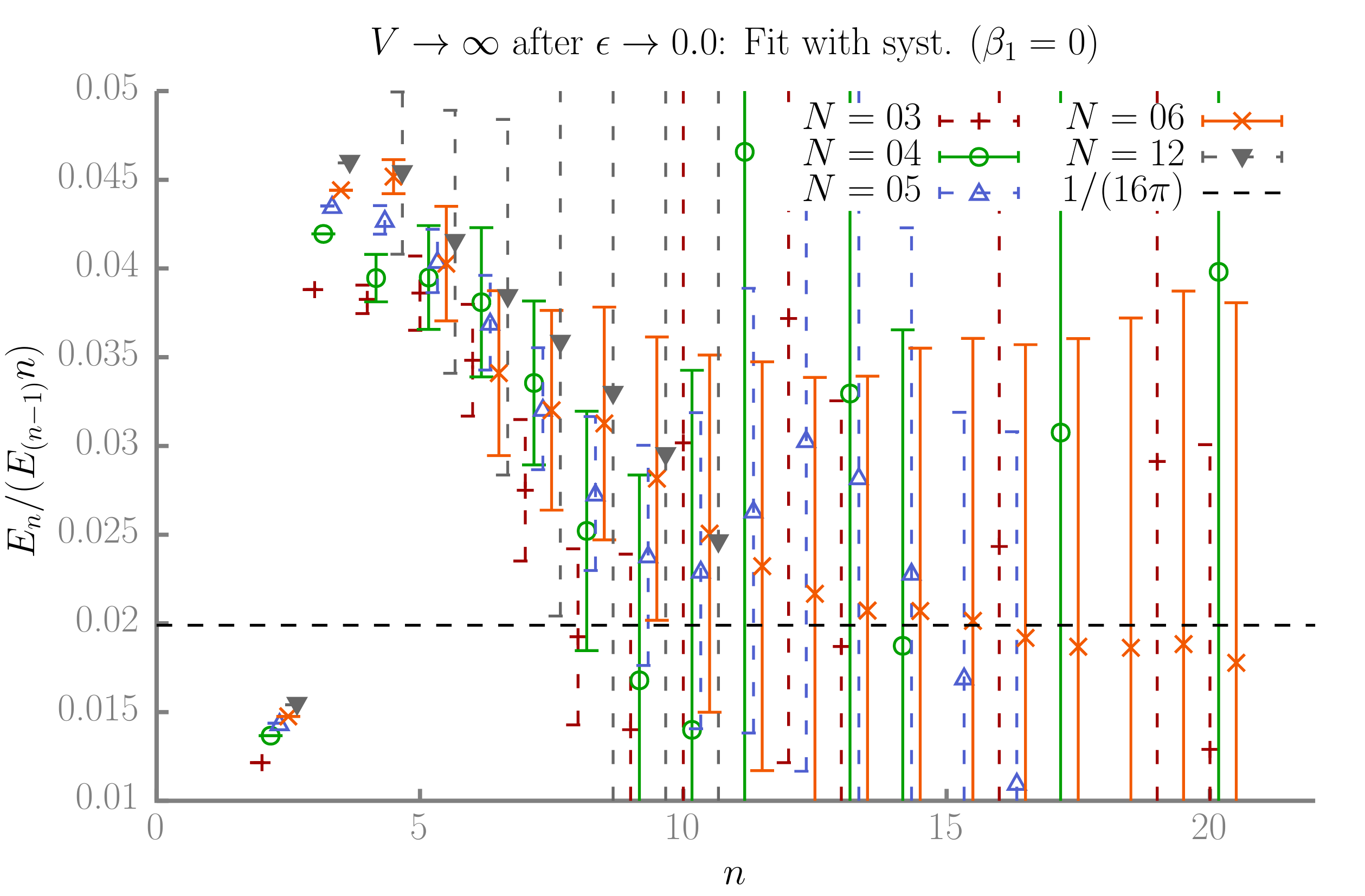}%
  \includegraphics[width=0.40\textwidth]{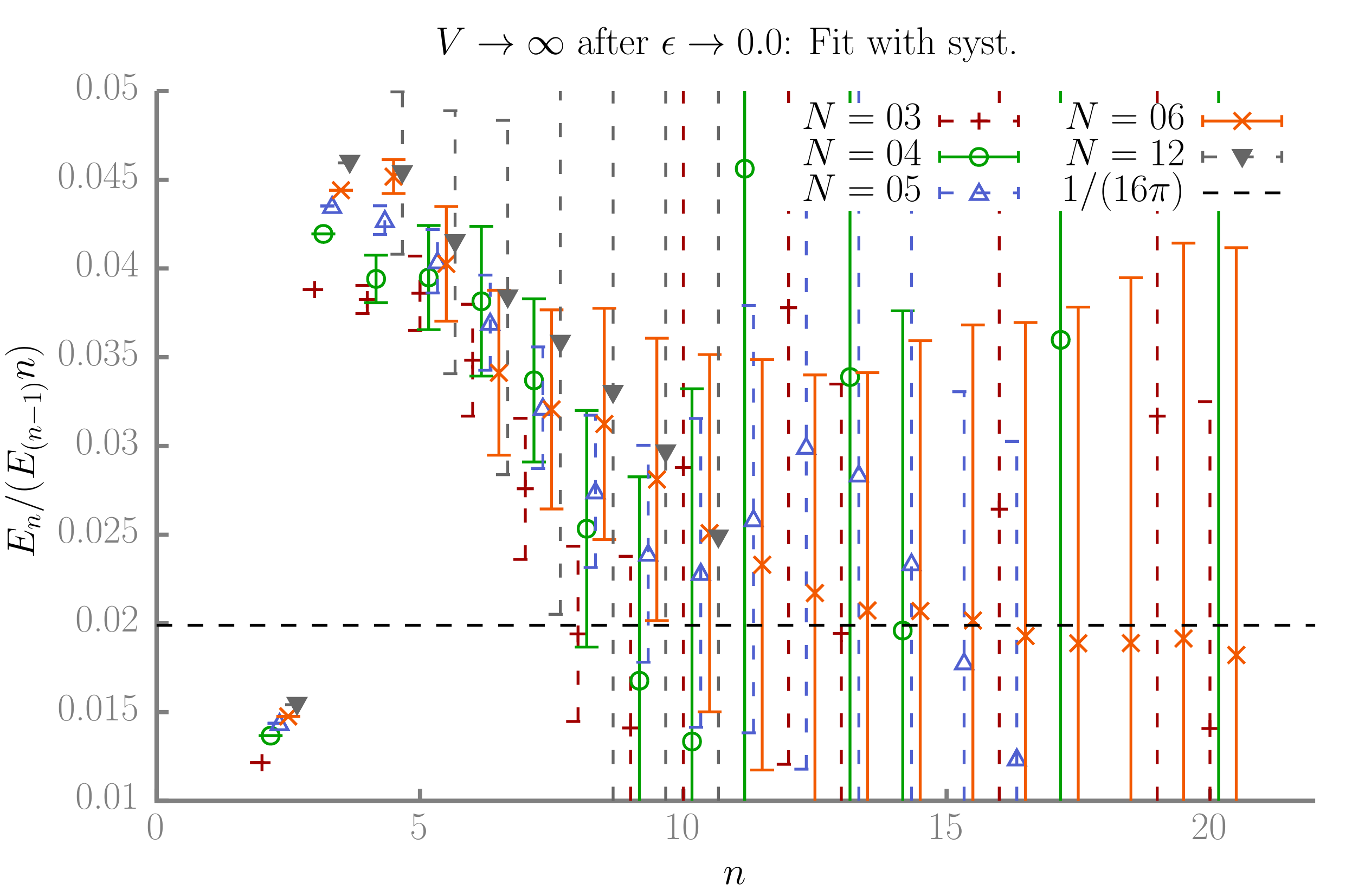}%
  \caption{Ratios after $\epsilon \to 0$ and $V \to \infty$ for fits including
      the coefficients $c_i$ [see Eq.~\eqref{eq:FV_expanison_2}].  In the left column, the
    $\beta$-function in the fit function is truncated after the $\beta_0$ term, the right
    column shows the fits with $\beta_0$ and $\beta_1$.  In the top row, systematic
    uncertainties for the $\dtau \to 0$ extrapolation are not taken into account, and in the
    bottom row they are estimated by Eq.~\eqref{eq:syst_error}.}
  \label{fig:rat_with_c}
\end{figure}

The plots in Fig.~\ref{fig:rat_with_c} clearly show that we get unacceptably large
uncertainties if the coefficients $c_i$ are included in our fits. More data and in
particular data for much larger lattice volumes would be needed to be able to capture the
difference between the $f_kc_i$ terms. 

\begin{figure}
  \centering
  \includegraphics[width=0.40\textwidth]{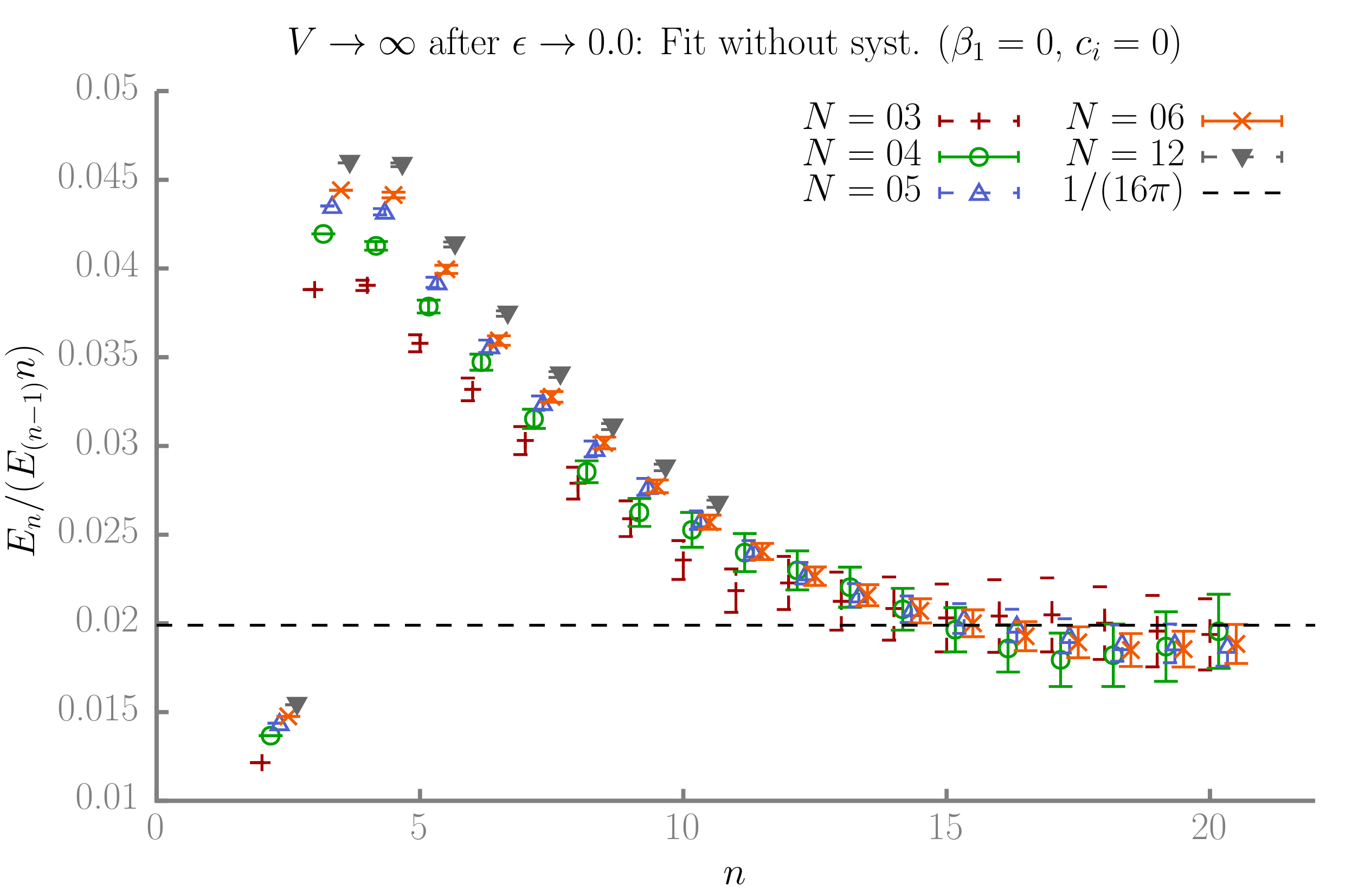}%
  \includegraphics[width=0.40\textwidth]{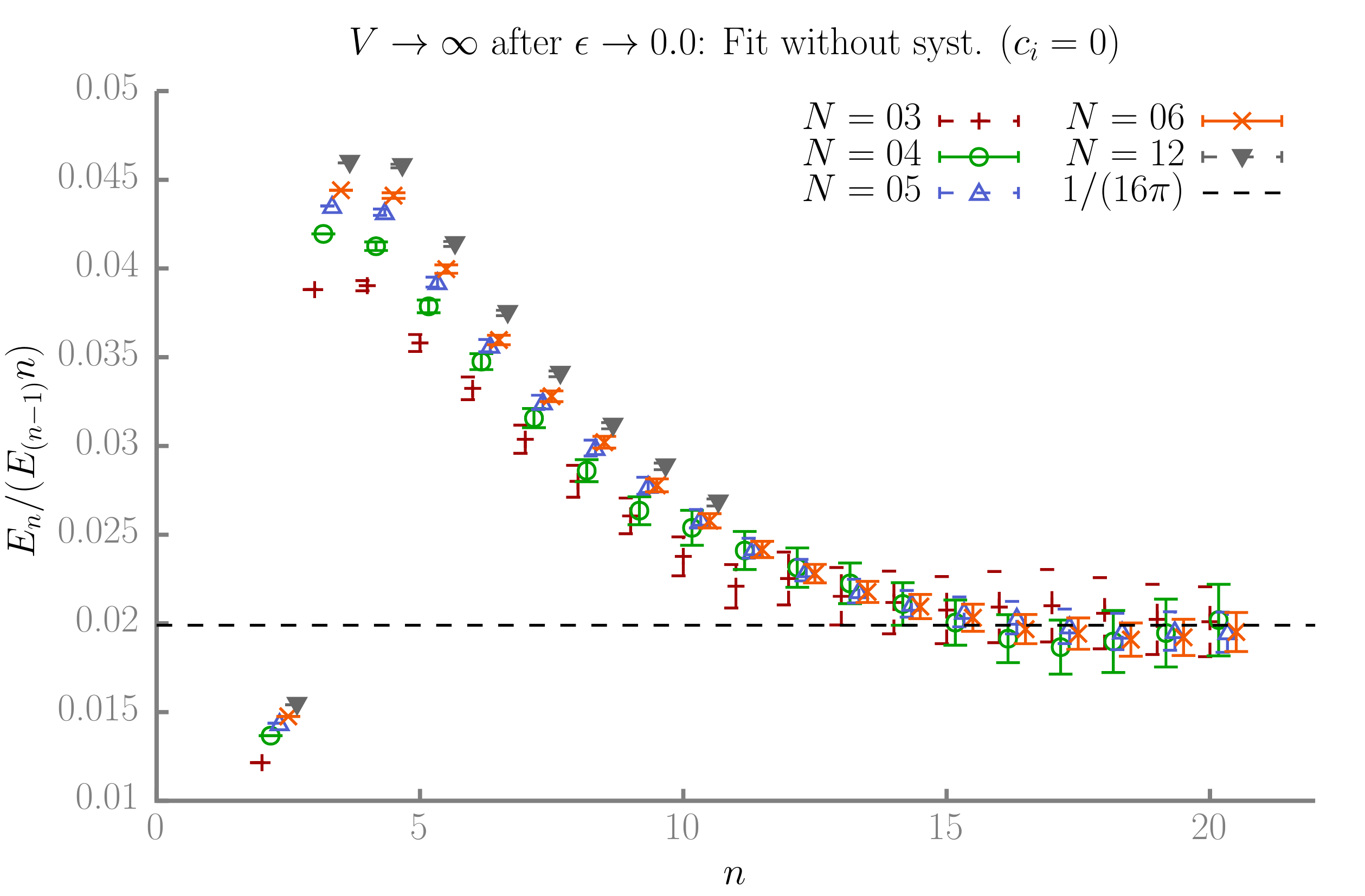} \\%
  \includegraphics[width=0.40\textwidth]{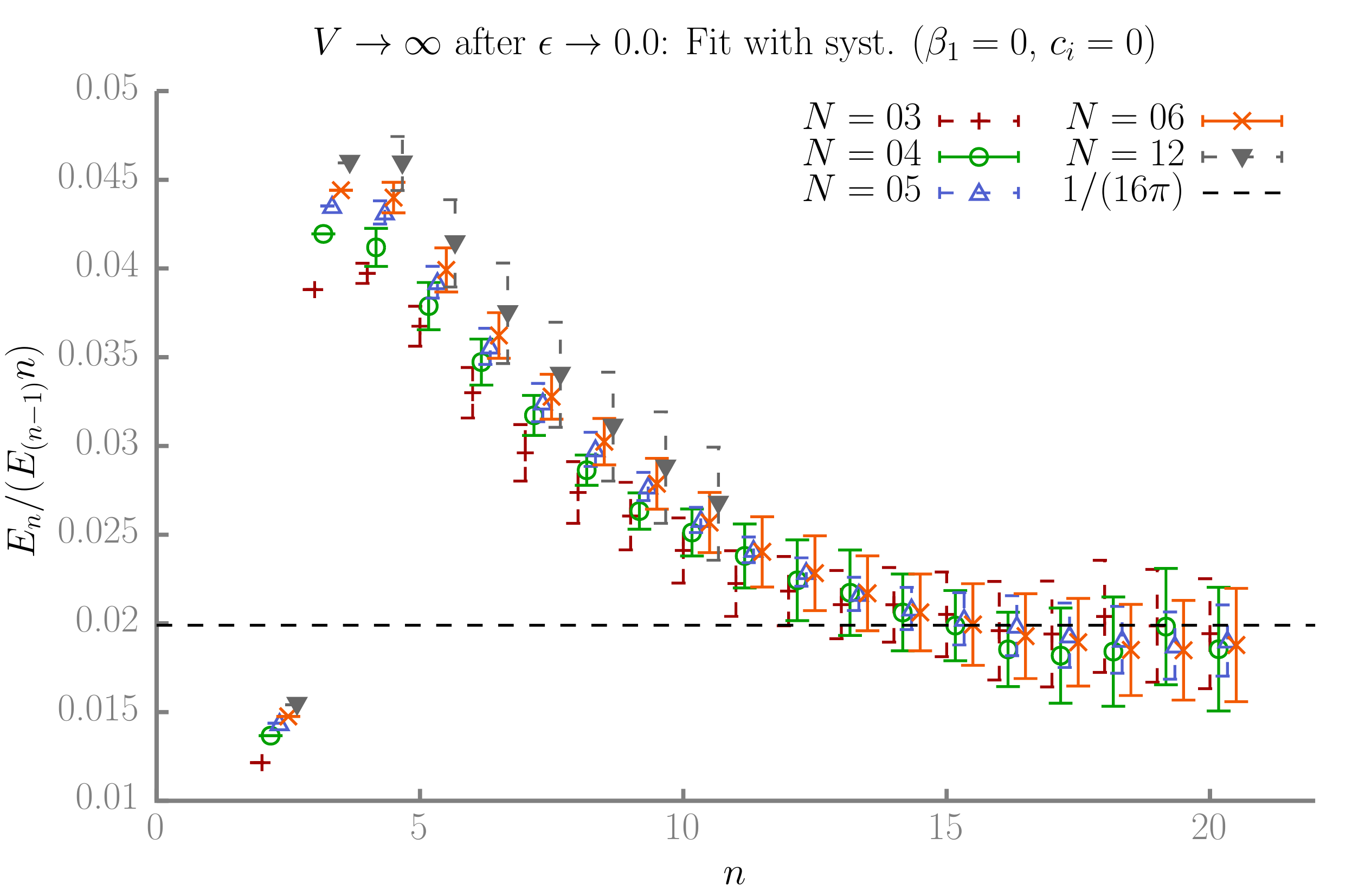}%
  \includegraphics[width=0.40\textwidth]{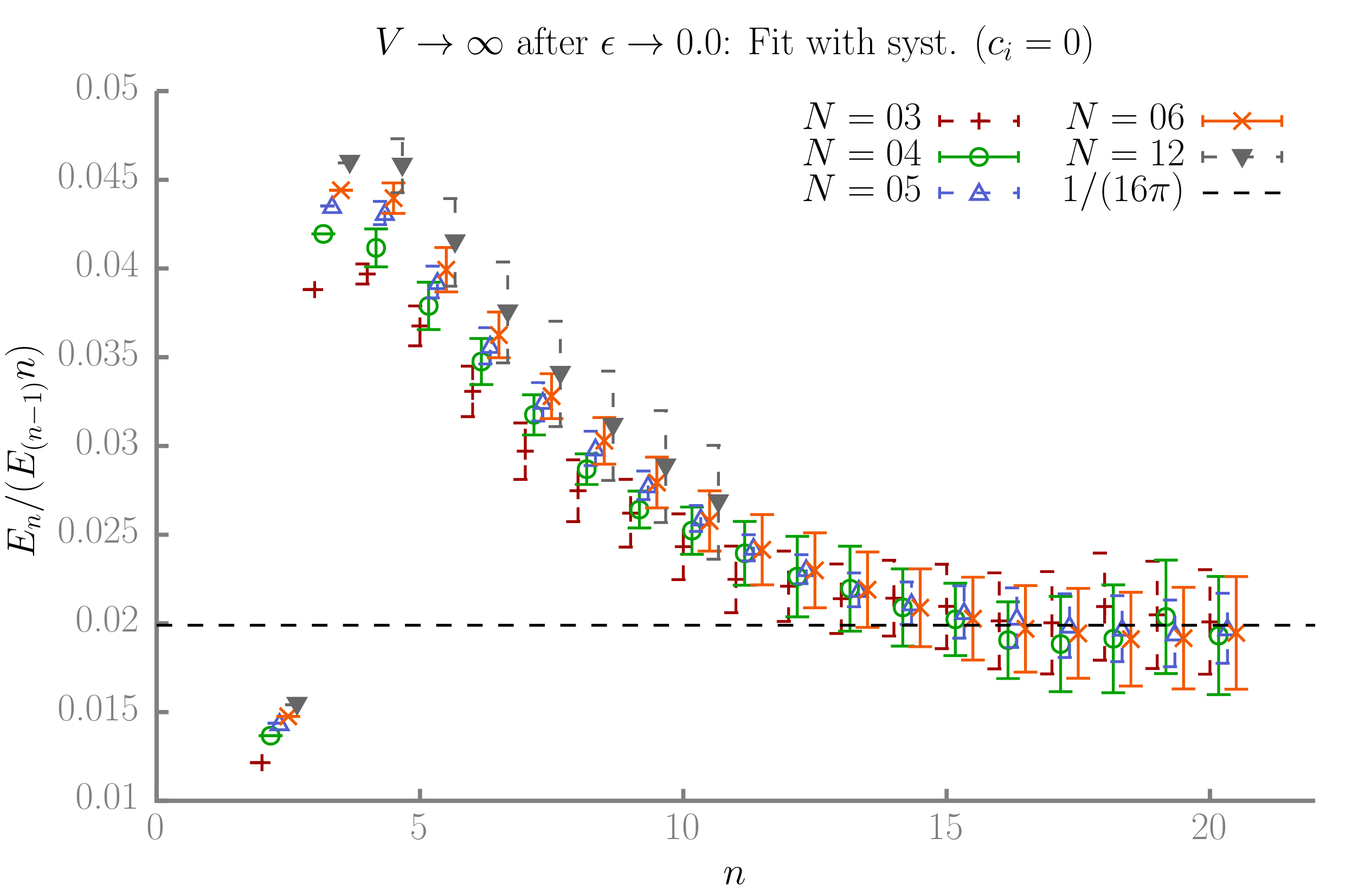}%
  \caption{Like Fig.~\ref{fig:rat_with_c} but with fit functions where the coefficients
    $c_i$ are set to zero. Within their respective uncertainties all the fits lead to
    indistinguishable results for the ratios $r_n$.}
  \label{fig:rat_no_c}
\end{figure}

Since we cannot resolve the $c_i$ anyhow, we leave them out of the fitting
procedure entirely. Setting $c_i=0$ drastically reduces the parameters in our fit
function, while keeping its general functional form --- a polynomial in $\ln(L)$ ---
intact. This works astonishingly well and the results are plotted in
Fig.~\ref{fig:rat_no_c}. Again, we find very small $\chi^2$ values of order $10^{-1}$ for
the fits with the systematic errors included. With the exception of the fits for $N=6$ and
$N=12$, even if we ignore the systematic errors the $\chi^2$ values are of order $1$.
All four fits yield, with overlapping error bars, the same final results for the
ratios. Moreover, the \emph{independent fits} for different $N$ lie on
top of each other after the asymptotic behavior sets in at $n \sim 15$.

\begin{figure}
  \centering
  \includegraphics[width=0.40\textwidth]{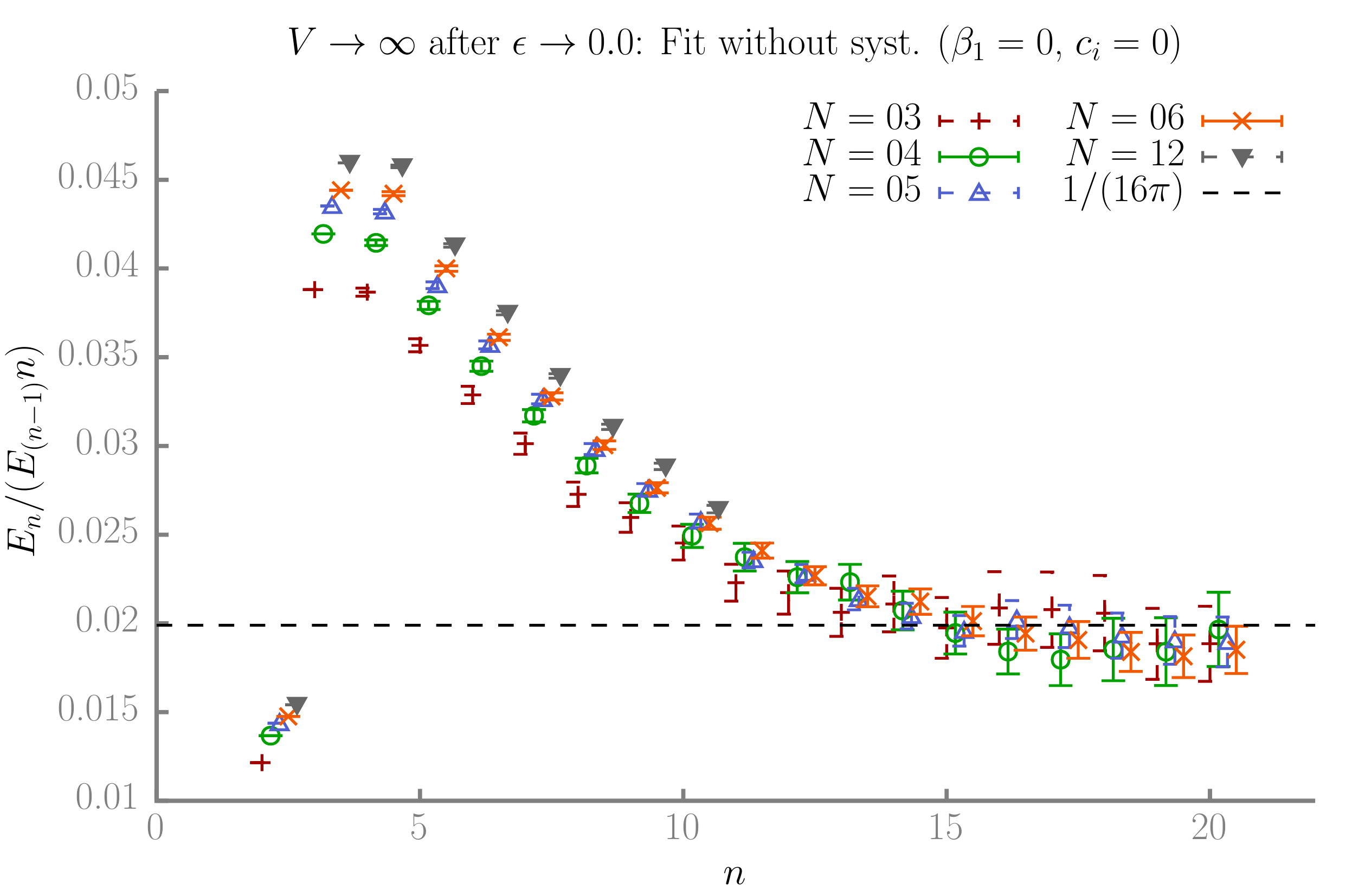}%
  \includegraphics[width=0.40\textwidth]{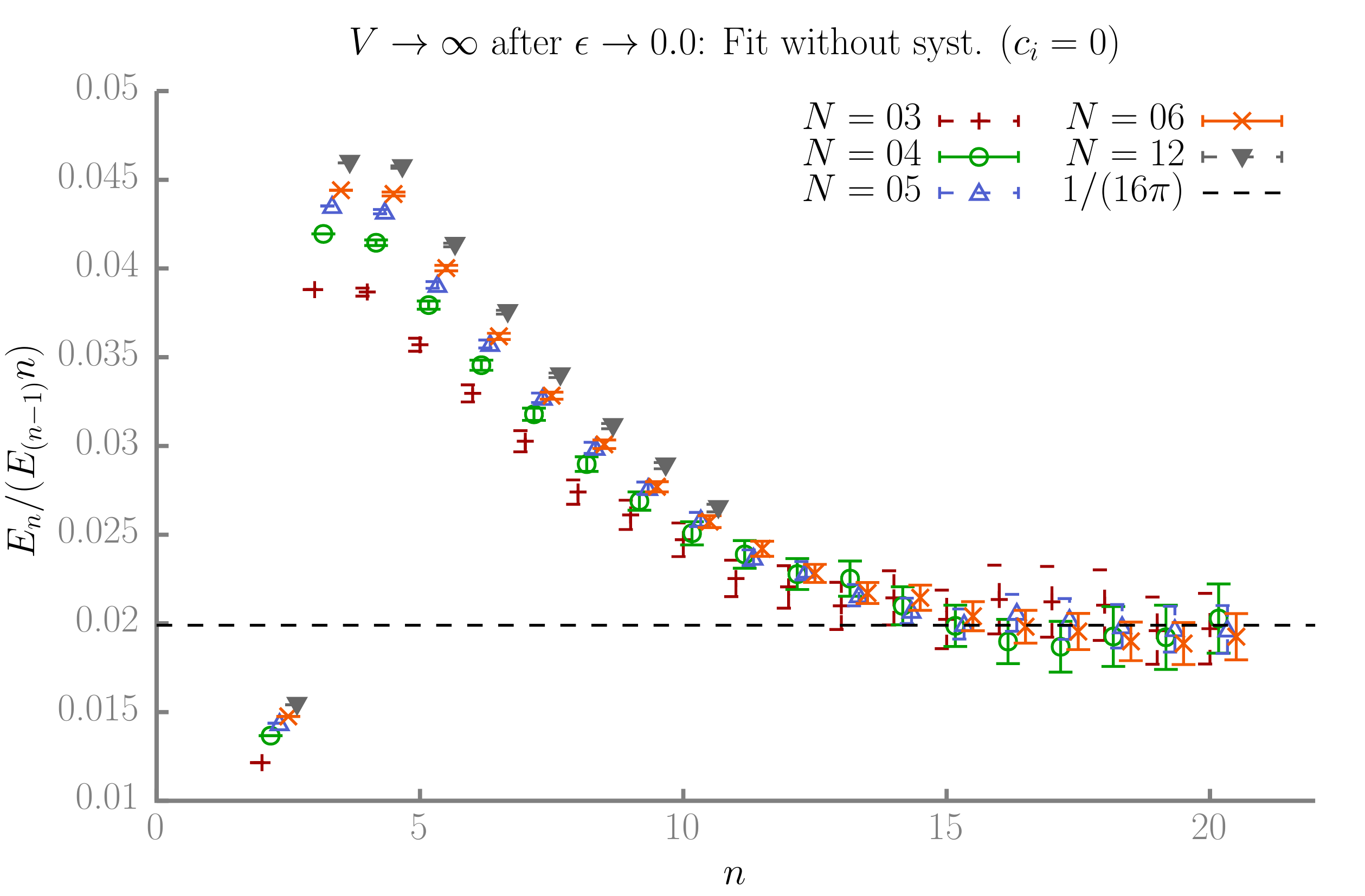} \\%
  \includegraphics[width=0.40\textwidth]{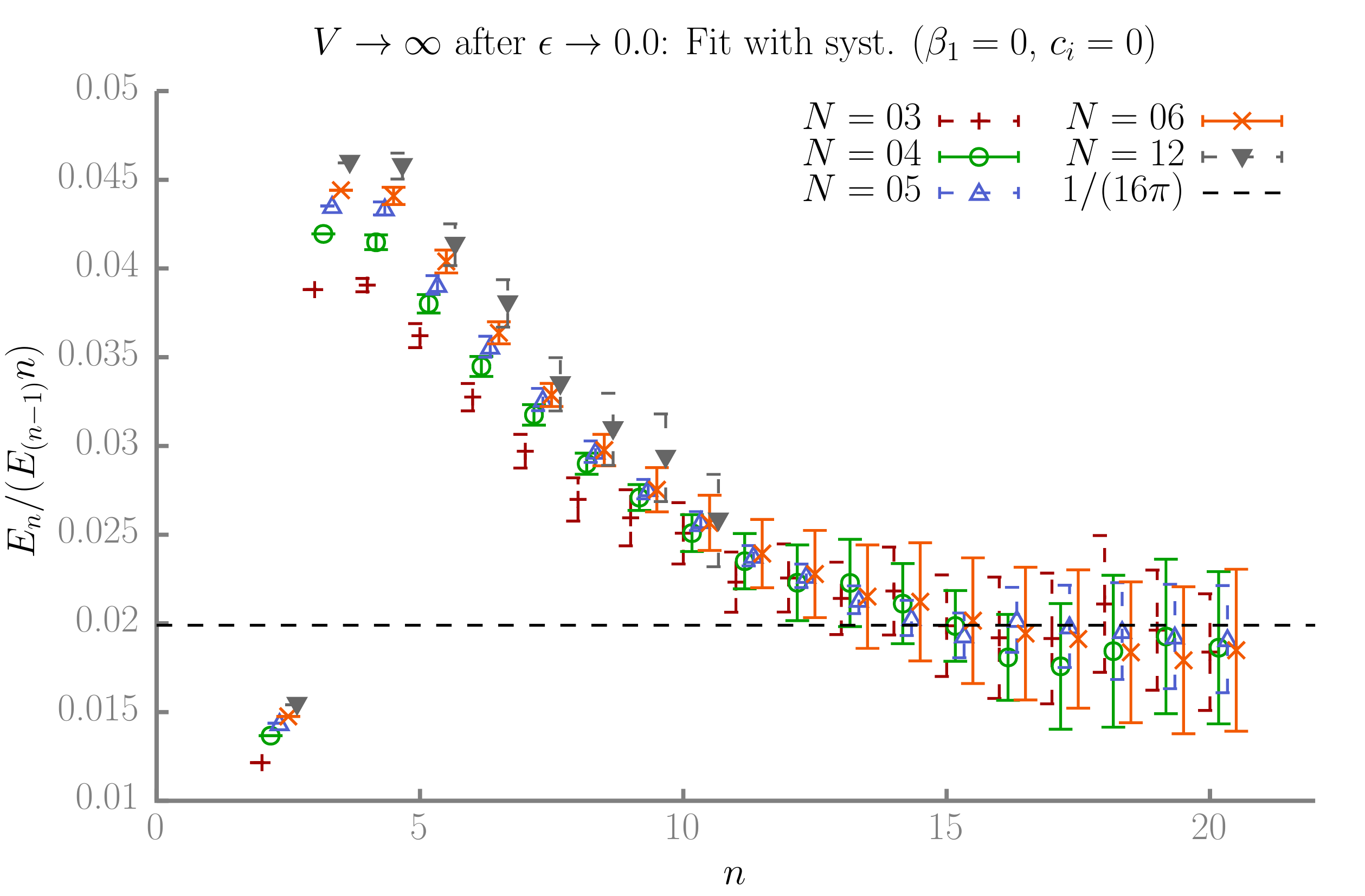}%
  \includegraphics[width=0.40\textwidth]{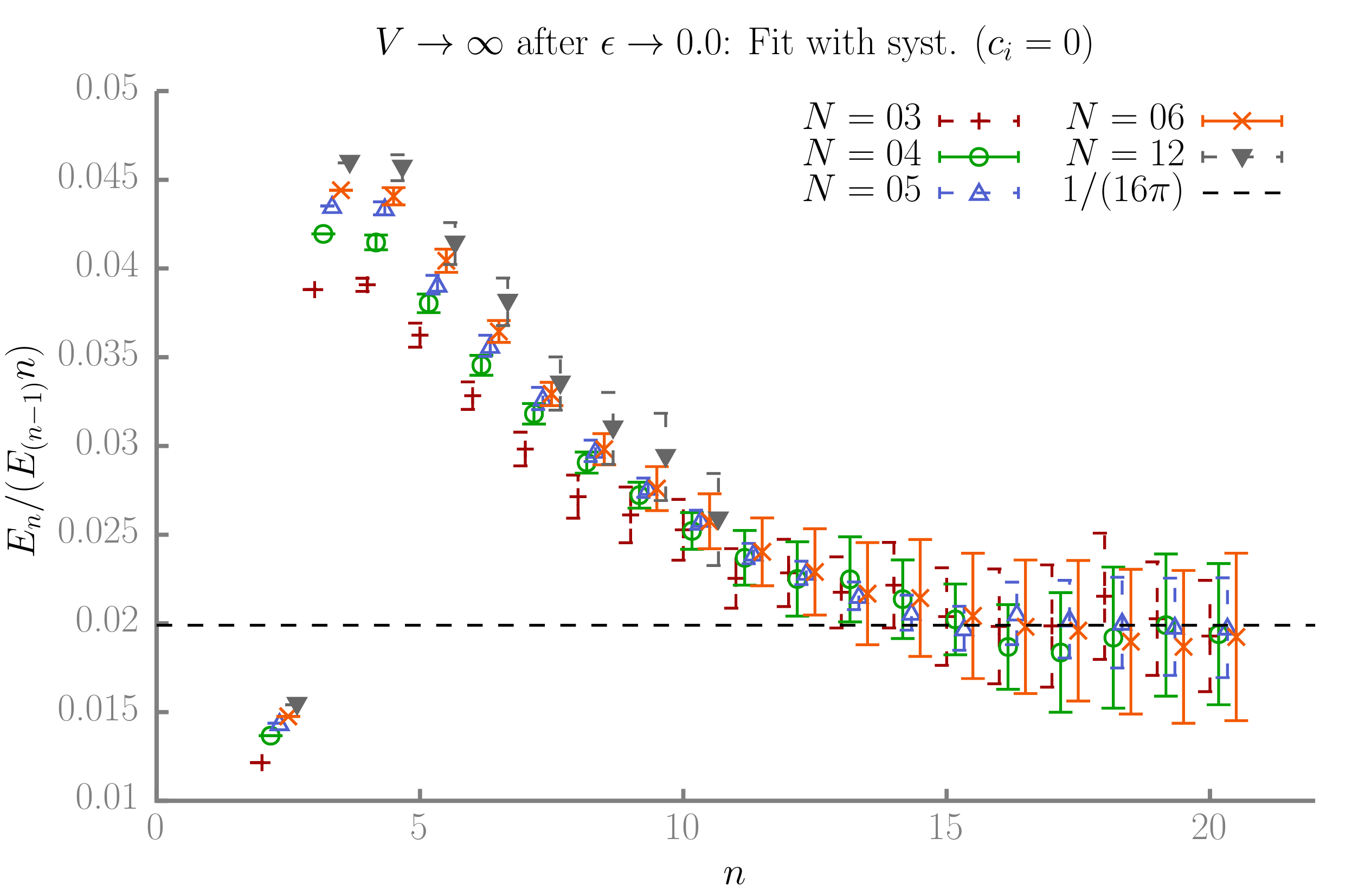}%
  \caption{Like Fig.~\ref{fig:rat_no_c} but for setup II.}
  \label{fig:rat_no_c_all}
\end{figure}

\begin{figure}
  \centering
  \includegraphics[width=0.40\textwidth]{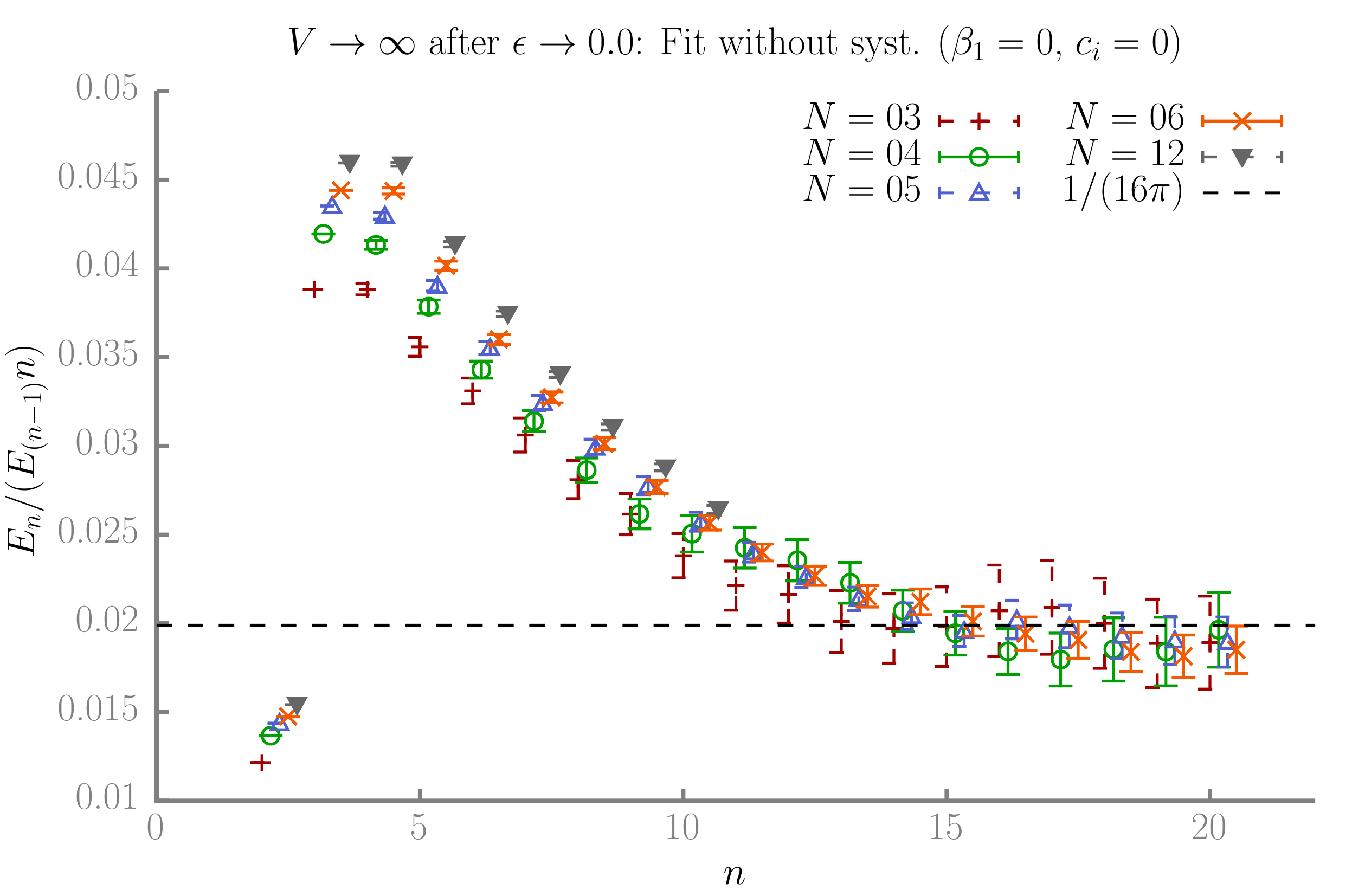}%
  \includegraphics[width=0.40\textwidth]{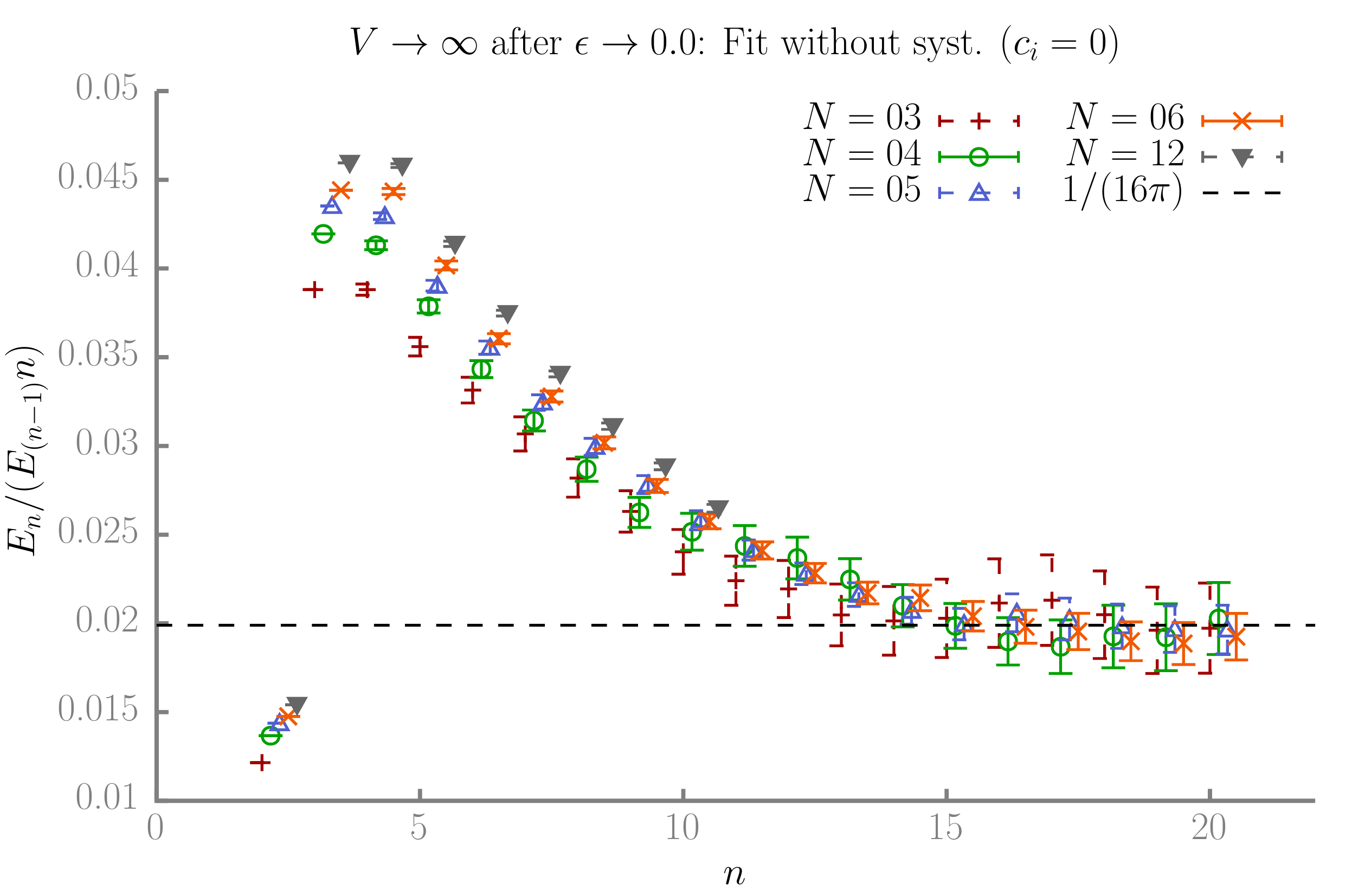} \\%
  \includegraphics[width=0.40\textwidth]{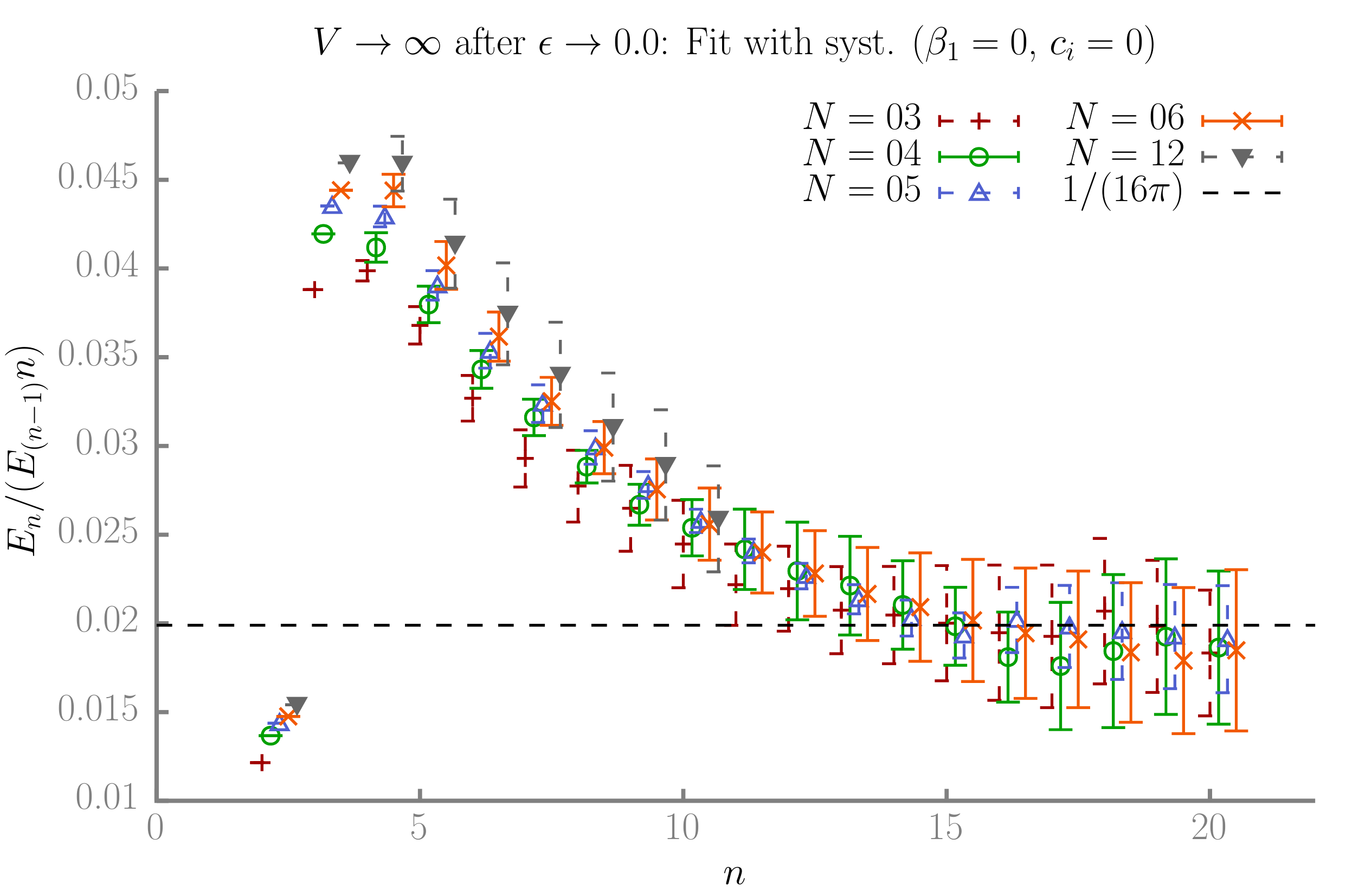}%
  \includegraphics[width=0.40\textwidth]{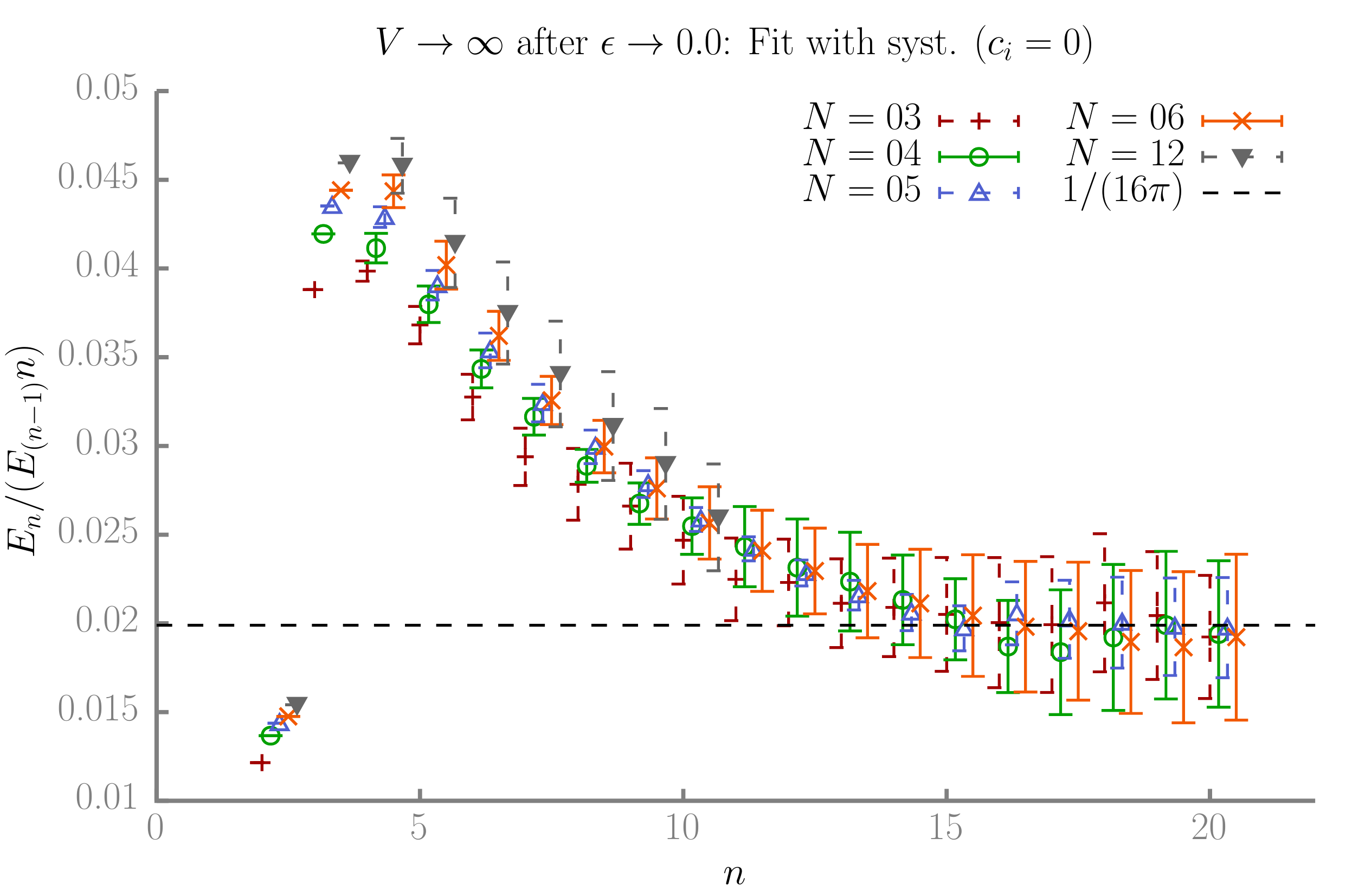}%
  \caption{Like Fig.~\ref{fig:rat_no_c} but for setup III. }
  \label{fig:rat_no_c_clean}
\end{figure}

\begin{figure}
  \centering
  \includegraphics[width=0.6\textwidth]{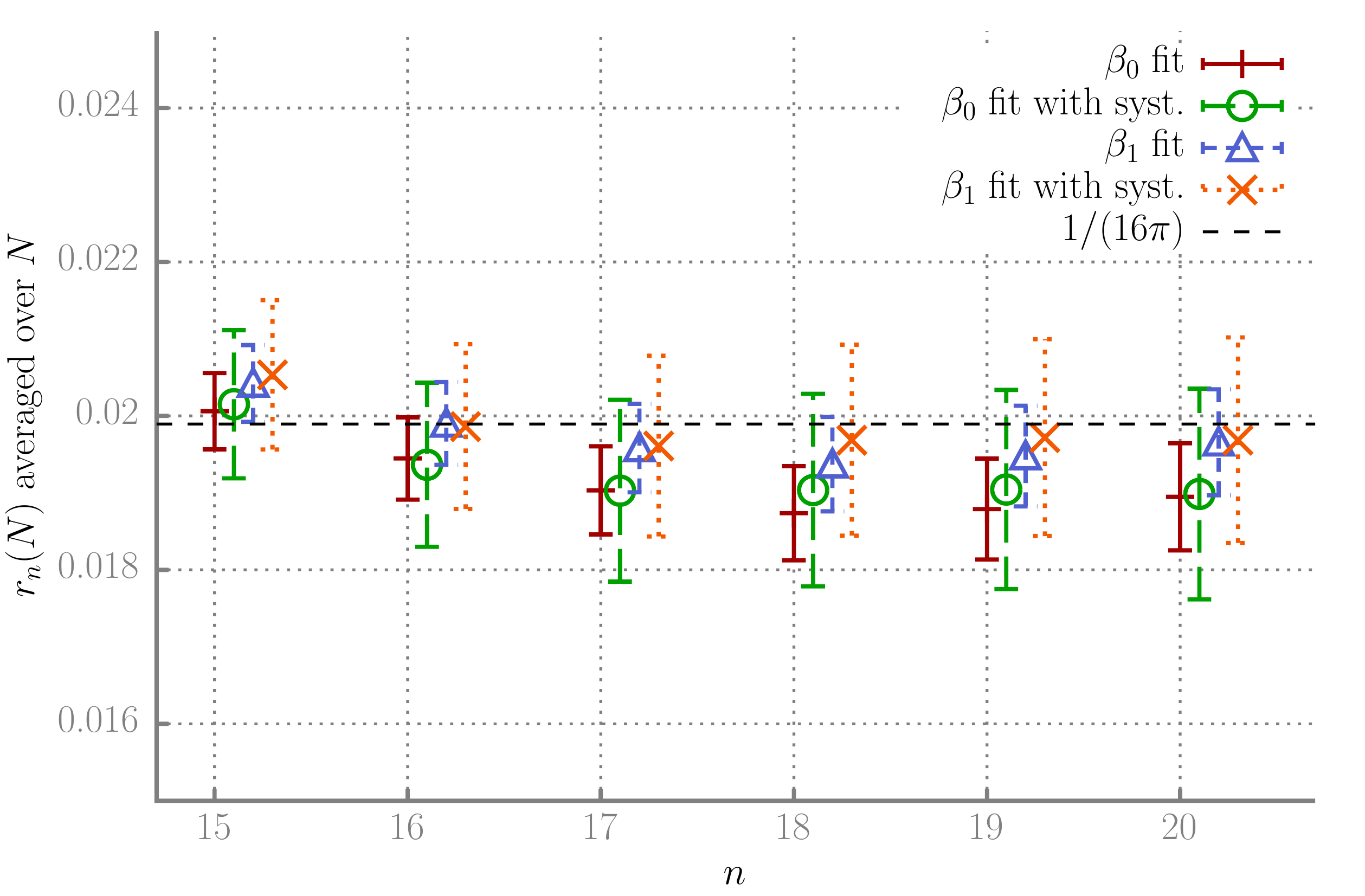} \\%
  \includegraphics[width=0.4\textwidth]{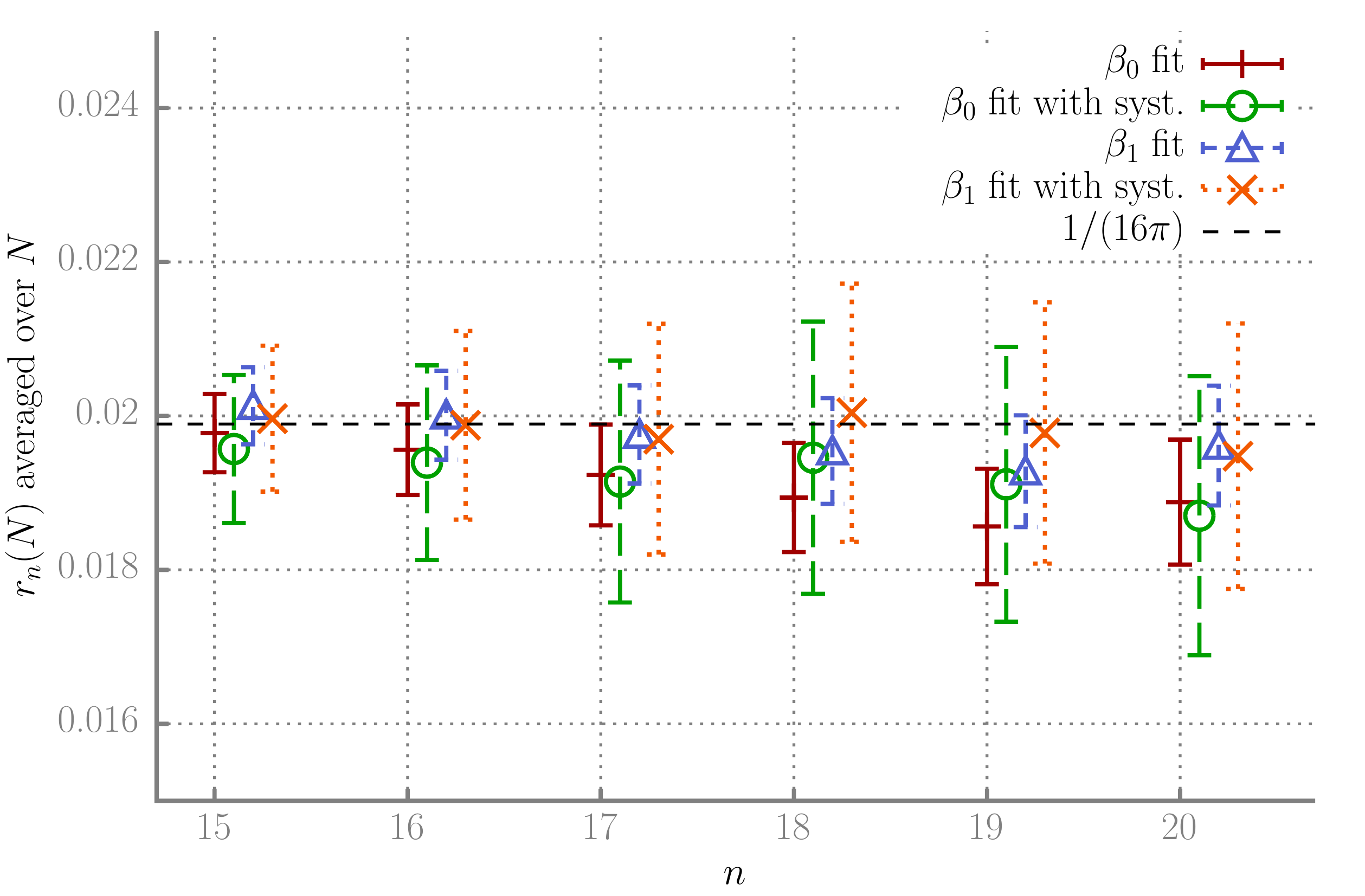}%
  \includegraphics[width=0.4\textwidth]{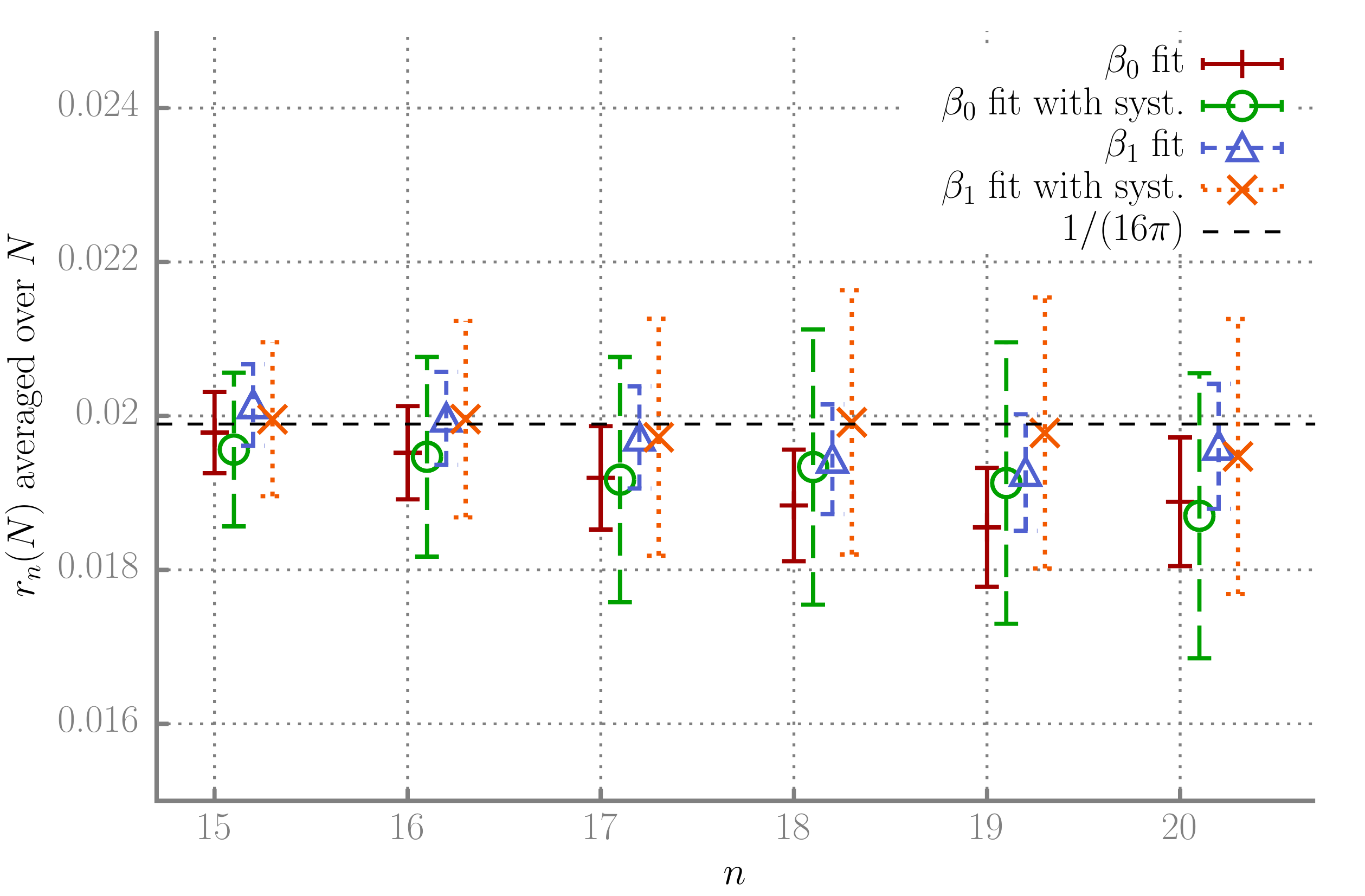}%
  \caption{Final results, constant fit to $r_n(N)$ for fixed $n$. Top: Data
  from setup~I. Bottom left: Data from setup~II. Bottom right: Data from setup~III.}
  \label{fig:NFit_all_datasets}
\end{figure}

The results plotted in Figs.~\ref{fig:rat_with_c} and \ref{fig:rat_no_c} are obtained
with the data from setup~I. It is essential to check how
strong the choice of the $\tau$ interval and the lattice configurations used for the fits
influence the final results. In Figs.~\ref{fig:rat_no_c_all} and
\ref{fig:rat_no_c_clean} we show the results for the ratios obtained with setup~II and
setup~III, respectively. Finally, in Fig.~\ref{fig:NFit_all_datasets} we show the final
results after averaging over the rank $N$ by fitting a constant to $r_n(N)$ for fixed $n$
for the setups and fit variants we considered. There is very good agreement between
all results.  We can therefore conclude that our extrapolation to infinite volume is very
robust and systematics from neglecting higher order terms in the beta function are
negligible. For the results in the main part of this work, we use the fits obtained with
finite $\beta_1$ and with systematic errors for the $\dtau \to 0$ extrapolation. We do not
include systematic error estimates for the truncation of the $\beta$-function, since
setting $\beta_1=0$ in the infinite volume extrapolation does not change the results
within uncertainties. 

\onecolumngrid

\twocolumngrid

\onecolumngrid

\twocolumngrid

\onecolumngrid

\twocolumngrid

\onecolumngrid

\twocolumngrid

\onecolumngrid

\twocolumngrid

\clearpage 


%

\clearpage

\end{document}